\documentclass[conference]{IEEEtran}
\IEEEoverridecommandlockouts
\usepackage{cite}
\usepackage{amsmath,amssymb,amsfonts}
\usepackage{graphicx}
\usepackage{textcomp}
\usepackage{xcolor}

\usepackage{tikz}
\usepackage{amsmath}
\usepackage{mathtools}
\usepackage{commath}
\usepackage{svg}
\usepackage{float}
\usepackage{bbm}
\usepackage{subfig}
\usepackage{filecontents}
\usepackage{amssymb}
\usepackage{bbm}
\usepackage[mathscr]{eucal}
\PassOptionsToPackage{hyphens}{url}\usepackage{hyperref}
\usepackage{url}
\usepackage{multirow}
\usepackage{subfig}
\usepackage{pifont}
\usepackage{enumitem}
\usepackage{diagbox}
\usepackage{xcolor}
\usepackage[linesnumbered,ruled,vlined]{algorithm2e}
\usepackage{ulem}
\usepackage{makecell}
\usepackage{tabularx}
\usepackage{cleveref}
\usepackage{verbatim}
\usepackage{url}

\usepackage{breakurl}

\DeclareMathOperator*{\argmax}{arg\,max}
\DeclareMathOperator*{\argmin}{arg\,min}

\def\BibTeX{{\rm B\kern-.05em{\sc i\kern-.025em b}\kern-.08em
    T\kern-.1667em\lower.7ex\hbox{E}\kern-.125emX}}
\begin{document}

\newcommand{\todo}[1]{{\textcolor{blue}{[ToDo: #1]}}}
\newcommand{\xmark}{\ding{55}}
\newcommand{\missing}{\ding{110}}

\newcommand\blfootnote[1]{%
  \begingroup
  \renewcommand\thefootnote{}\footnote{#1}%
  \addtocounter{footnote}{-1}%
  \endgroup
}

\newcommand{\fk}[3]{%
\ifx\hfuzz#1\hfuzz%
\else%
\sout{#1} %
\fi%
#2%
\ifx\hfuzz#3\hfuzz%
\else%
{ \textcolor{red}{[fk: #3]}}%
\fi%
}
\newcommand{\ej}[1]{{\textcolor{magenta}{[ej: #1]}}}
\newcommand{\xl}[1]{{\textcolor{brown}{[xl: #1]}}}
\newcommand{\nils}[1]{{\textcolor{orange}{[nl: #1]}}}
\title{SoK: How Robust is Image Classification Deep Neural Network Watermarking?\\ (Extended Version)}

\author{
\IEEEauthorblockN{Nils Lukas, Edward Jiang, Xinda Li, Florian Kerschbaum}
\IEEEauthorblockA{\textit{University of Waterloo}\\
Waterloo, Canada \\
\{nlukas, eydjiang, xinda.li, florian.kerschbaum\}@uwaterloo.ca}
}

\maketitle

\begin{abstract}
Deep Neural Network (DNN) watermarking is a method for provenance verification of DNN models. 
Watermarking should be robust against watermark removal attacks that derive a \emph{surrogate} model that evades provenance verification.
Many watermarking schemes that claim robustness have been proposed, but their robustness is only validated in isolation against a relatively small set of attacks.
There is no systematic, empirical evaluation of these claims against a common, comprehensive set of removal attacks. 
This uncertainty about a watermarking scheme's robustness causes difficulty to trust their deployment in practice.
In this paper, we evaluate whether recently proposed watermarking schemes that claim robustness are robust against a large set of removal attacks. 
We survey methods from the literature that (i) are known removal attacks, (ii) derive surrogate models but have not been evaluated as removal attacks, and (iii) novel removal attacks. 
\emph{Weight shifting} and \emph{smooth retraining} are novel removal attacks adapted to the DNN watermarking schemes surveyed in this paper.
We propose taxonomies for watermarking schemes and removal attacks. 
Our empirical evaluation includes an ablation study over sets of parameters for each attack and watermarking scheme on the image classification datasets CIFAR-10 and ImageNet. 
Surprisingly, our study shows that none of the surveyed watermarking schemes is robust in practice. 
We find that schemes fail to withstand adaptive attacks and known methods for deriving surrogate models that have not been evaluated as removal attacks.
This points to intrinsic flaws in how robustness is currently evaluated. 
Our evaluation includes a discussion of the runtime of each attack to underpin their practical relevance. 
While none of the schemes is robust against all attacks, none of the attacks removes all watermarks. 
We show that attacks can be combined and find combined attacks that remove all watermarks. 
We show that watermarking schemes need to be evaluated against a more extensive set of removal attacks with a more realistic adversary model.
Our source code and a complete dataset of evaluation results are publicly available, which allows to independently verify our conclusions.
\end{abstract}


\section{Introduction}

\blfootnote{A shorter version of this paper~\cite{real_paper} is to appear at IEEE S\&P'22. }
Deep Neural Networks (DNN) have become state-of-the-art algorithms for applications such as facial recognition~\cite{parkhi2015deep, liu2017sphereface, wang2018cosface}, medical image classification~\cite{zhang2019medical} and autonomous driving~\cite{luo2017traffic}. 
Training a DNN model can be expensive due to data preparation (collection, organizing, and cleaning) and computational resources required for validating a model~\cite{press2016}. 
For this reason, DNNs are often provided by a single entity and consumed by many, such as in Machine Learning-as-a-Service (MLaaS). 
A model provider may want to restrict unauthorized redistribution of their \emph{source} model. 
The threat to the model provider is a user who derives a (stolen) \emph{surrogate} model from access to the source model and publicly deploys their surrogate model. 
Krishna et al.~\cite{krishna2020thieves} have shown that such \emph{model stealing} attacks can be (i) effective because even high-fidelity surrogates of large models like BERT~\cite{devlin2018bert} can be derived with limited access to domain data and (ii) practical because surrogate models can be derived for a fraction of the costs compared to retraining a model. 

Papernot et al.~\cite{papernot2018sok} describe the \emph{confidentiality} requirement as one of the core principles for security and privacy in machine learning. 
Preserving a model's confidentiality refers to protecting its parameters against model stealing attacks.
Confidentiality is important because the source model constitutes intellectual property and may leak information about its training dataset. 
Preventing model stealing is difficult~\cite{jagielski2020high, carlini2020cryptanalytic, krishna2020thieves, atli2020extraction}, but detecting whether the confidentiality of a source model has been broken serves as a powerful deterrent and can be achieved through DNN \emph{watermarking}. 

DNN watermarking~\cite{uchida2017embedding} is a method designed to detect surrogate models. 
Watermarking embeds a message into a model that is later extractable using a secret key.
Developing DNN watermarking schemes is an active area of research studied by large corporations such as Microsoft~\cite{rouhani2018deepsigns}, Google~\cite{adi2018turning} and IBM~\cite{zhang2018protecting}. 
Robustness is a core security property of watermarking, which states that an attacker cannot derive surrogate models from access to the source model that do not retain the watermark. 
Watermarking schemes that are robust against such \emph{watermark removal} attacks are needed to deter redistribution by adversaries. 
Claimed security properties of some existing watermarking schemes~\cite{adi2018turning, zhang2018protecting} had been broken by novel attacks~\cite{shafieinejad2019robustness, liu2020removing, wang2019attacks}, but it is unclear how these attacks generalize to other watermarks. 

We perform a systematic evaluation and propose taxonomies for watermarking schemes and attacks.
We survey 29 methods from the literature that (i) are known removal attacks, such as weight pruning~\cite{zhu2017prune} or knowledge distillation~\cite{hinton2015distilling}, (ii) derive surrogate models but have not been evaluated as removal attacks, and (iii) novel removal attacks. 
A removal attack is \emph{effective} if the surrogate model has a high test accuracy and does not retain the watermark. 
It is \emph{efficient} if resources required to run the attack, such as its runtime, are small compared to retraining a model from scratch.
We measure both effectiveness and efficiency. 
In our taxonomy, we categorize attacks into (i) model modification, (ii) input preprocessing, and (iii) model extraction. 
Model modification and input preprocessing modify the source model or its input, whereas model extraction trains a different surrogate model by distilling knowledge from the source model. 

We survey eleven\footnote{Zhang et al.~\cite{zhang2018protecting} propose three different schemes.} recently proposed watermarking schemes~\cite{uchida2017embedding, adi2018turning, chen2018deepmarks, chen2019blackmarks, jia2020entangled, le2020adversarial, rouhani2018deepsigns,szyller2019dawn,zhang2018protecting} from the literature that claim robustness.
Most of these schemes do not specify whether their definition of robustness includes model extraction~\cite{zhang2018protecting, uchida2017embedding, rouhani2018deepsigns, chen2018deepmarks, chen2019blackmarks}, one scheme restricts the runtime of the attacker~\cite{adi2018turning} and the remaining schemes claim robustness against any removal attack~\cite{le2020adversarial, jia2020entangled, szyller2019dawn}. 
In this paper, we evaluate robustness against any removal attack and demonstrate whether an attack is efficient by showing its runtime. 
Our taxonomy categorizes these watermarking schemes into (i) model independent, (ii) model dependent, (iii) parameter encoding, and (iv) active watermarking schemes. 

Our new Watermark-Robustness-Toolbox (WRT) implements all watermarking schemes and removal attacks evaluated in this paper.
We validate the robustness of each scheme against each removal attack. 
Our evaluation includes an ablation study over multiple sets of parameters for each watermarking scheme and removal attack.
The defender and attacker engage in a zero-sum game to choose the best parameter set for their method, which constitutes the Nash equilibrium.
We say a scheme is robust if the defender can choose a set of parameters so that no removal attack is effective. 
Our study analyzes the robustness of watermarking schemes and the effectiveness and efficiency of removal attacks. 
We also study the robustness of watermarking scheme categories against categories of removal attacks to identify the category of most effective attacks that should be used to evaluate the robustness of a watermarking scheme in a specific category. 

Our empirical evaluations are performed on large datasets to emphasize the practical relevance of our work. 
The experiments span CIFAR-10~\cite{cifar10} and ImageNet~\cite{imagenet}, which are image classification datasets.  
The ImageNet dataset contains over 1.2 million training images from 1k categories and is a broadly accepted benchmark to measure the performance of state-of-the-art machine learning models~\cite{radford2learning}. 

Our study shows that none of the investigated watermarking schemes is robust against all removal attacks. 
However, we also find that none of the attacks from the literature removes all watermarks. 
We propose new \emph{combined} attacks that remove all investigated watermarks while maintaining a high test accuracy in the surrogate models.
Our study also shows that robustness should be verified against a more extensive set of attacks and on a larger number of datasets.  
We believe that an open-source implementation of watermarking schemes and removal attacks enhances the scientific study of a scheme's robustness. 
Towards this goal, we make our new Watermark-Robustness-Toolbox (WRT) and a complete dataset of evaluation results publicly available with documentation, which allows independently verifying our conclusions.

\begin{table}[]
    \centering
    \begin{tabular}{|c|c|}
        \hline
         \textbf{Requirements} & \textbf{Description} \\ \hline
         Fidelity & \makecell{The impact on the model's task accuracy is small. }\\ \hline
         Robustness & \makecell{Surrogate models retain the watermark. } \\ \hline
         Integrity & \makecell{Models trained without access to the source model\\ do not retain the watermark.}\\ \hline
         Capacity & \makecell{The watermark allows encoding large messages sizes.  }\\ \hline
         Efficiency & \makecell{Embedding and extracting the watermark is efficient. }\\ \hline
         Undetectability & \makecell{The watermark cannot be detected efficiently\\ without knowledge of the secret watermarking key. }\\ \hline
    \end{tabular}
    \caption{Requirements for ideal DNN watermarking.}
    \label{tab:requirements}
\end{table}

\subsection{Contributions}
This work contributes:
\begin{itemize}
    \item Taxonomies of DNN watermarking schemes and removal attacks.
    \item An empirical evaluation of the robustness of DNN watermarking schemes~\cite{uchida2017embedding, adi2018turning, chen2018deepmarks, chen2019blackmarks, jia2020entangled, le2020adversarial, rouhani2018deepsigns,szyller2019dawn,zhang2018protecting} against removal attacks from related work.
    \item A unified adversary model for the attacker and defender in any of the evaluated watermarking schemes. 
    \item Proposal of the novel removal attacks \emph{weight shifting} and \emph{smooth retraining}.
    \item Combined attacks that remove all surveyed watermarks.
    \item Guidelines to evaluate the robustness of watermarking. 
    \item An open-source implementation of all watermarking schemes and removal attacks evaluated in this paper. 
\end{itemize}

\subsection{Organization}
The rest of the paper is organized as follows.
Section~\ref{sec:background} describes background information on deep neural networks.  
Section~\ref{sec:taxonomy} presents our taxonomy on watermarking schemes and removal attacks and
Section~\ref{sec:adversary_model} describes a unified adversary model for the attacker and defender. 
Section~\ref{sec:methodology} presents the methodology for our experiments and defines all measured quantities for our experiments. 
Empirical results are presented in Section~\ref{sec:experiments}.
Section~\ref{sec:guidelines} presents guidelines for evaluating robustness and Section~\ref{sec:conclusion} concludes the paper. 
Descriptions of the watermarking schemes and attacks and parameters for our ablation study can be found in Appendix~\ref{sec:schemes} and \ref{sec:attacks}. 

\section{Background}
\label{sec:background}

\subsection{Deep Neural Networks (DNNs)}
A deep neural network (DNN) classifier is a function $M: \mathcal{X} \rightarrow \mathcal{Y}$ that assigns a likelihood to inputs $\mathcal{X}\subseteq \mathbb{R}^d$ for each of $K \in \mathbb{N}$ classes $\mathcal{Y} \subseteq \mathbb{R}^K$. It is a sequence of layers $f_i, (i\in\{1,..,L\})$ in which each layer implements a linear function followed by a non-linear function called the activation function. 
A neural network is called deep if it has more than one layer between the input and output layer, called hidden layers. Hidden layers have weight and bias parameters used to compute that layer's activations. 
A softmax activation function $\sigma(\cdot)$ is applied to the output layer $f_L(\cdot)$ to convert likelihoods into probabilities for each predicted class. 
\begin{align}
    \sigma(f_L(x))_i=\frac{\exp(f_L(x)_i)}{\sum_j \exp(f_L(x)_j)}
\end{align}
Training a neural network model requires the specification of a differentiable loss function that is optimized by gradient descent on all trainable weights and biases. One such loss function is the cross-entropy loss $H$ for some ground truth $y\in \mathcal{Y}$ with respect to the model's prediction.
\begin{align}
    H(y, f_L(x)) = -\sum_{0 \leq k < K }(y_k \cdot \log(\sigma(f_L(x))_k))
\end{align}
A black-box deployment of a DNN exposes only the API of the model. 
On input of an element $x\in \mathcal{X}$, the server responds  with the full confidence vector $\sigma(f_L(x))\in \mathcal{Y}$.

\section{Taxonomy of Watermarking}
\label{sec:taxonomy}

In this section, we define DNN watermarking and describe our proposed taxonomy. 
We introduce watermarking as a method for DNN provenance verification and propose categorizations of watermarking schemes and removal attacks. 
\begin{figure}
    \includegraphics[width=1.\linewidth]{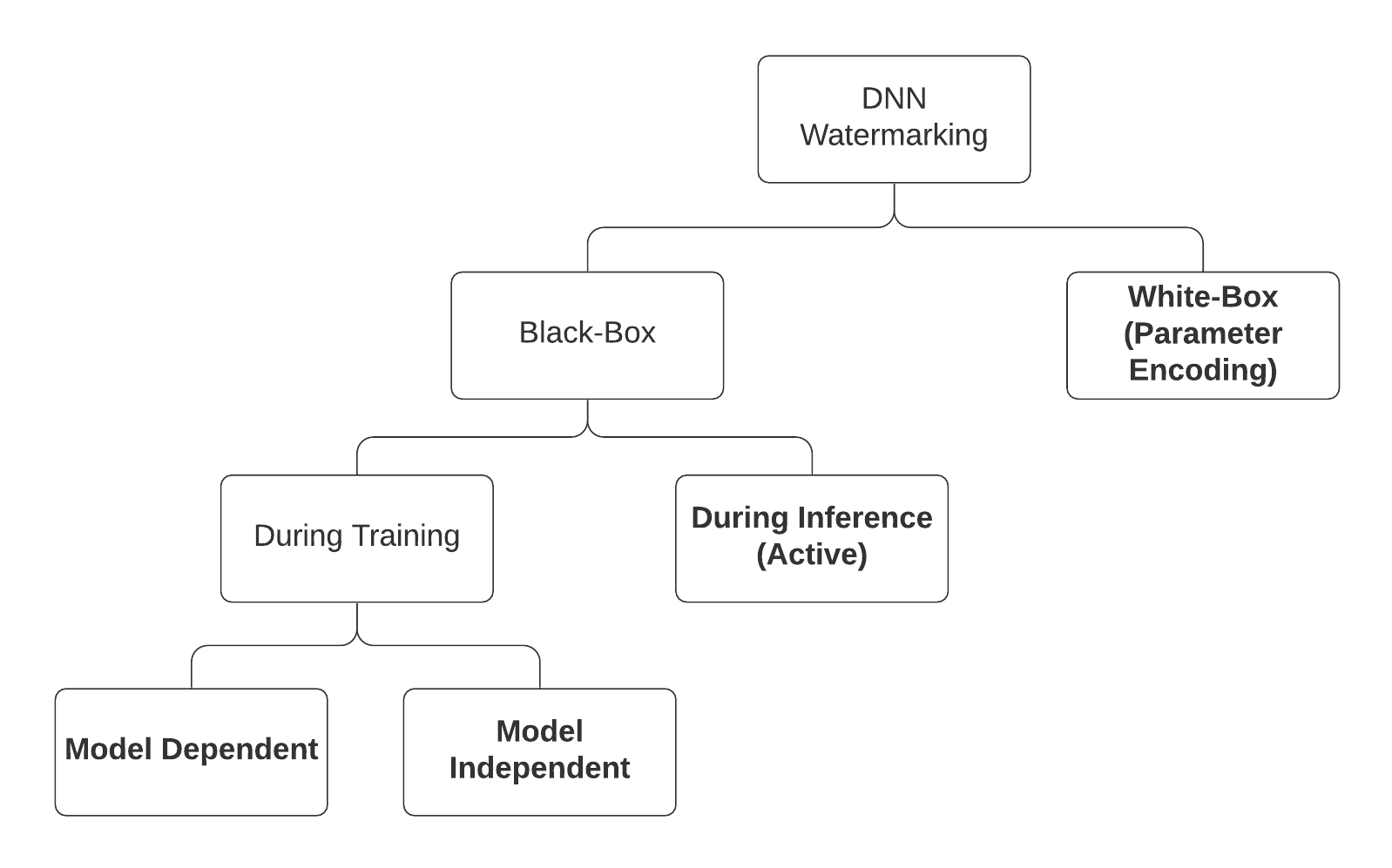}
    \caption{A categorization of watermarking schemes. The distinction between 'white-box' and 'black-box' refers to the verification requirement, whereas 'during training' and 'during inference' refer to the embedding of the watermark. }
    \label{fig:tree_diagram}
\end{figure}

\subsection{Defining Watermarking.}
\label{sec:defining_watermarking}
Watermarking embeds a message into a \emph{source model} that is later extractable using a secret watermarking key.
The \emph{success rate} between two messages can be computed as the number of matching bits normalized by the message length. 
It is defined as follows for messages $a, b \in \{0,1\}^N$ of size $N$, where $\delta$ denotes the Kronecker delta. 
\begin{align*}
    \Delta(a, b) = \frac{1}{N}\sum_{i=1..|N|} \delta(a_{i}, b_{i})
\end{align*}
A watermark is \emph{retained} in a model if the same message can be extracted with a success rate that is higher than a \emph{decision threshold}, defined by the watermarking scheme. 
Otherwise, we say that a watermark is removed.
A watermark should be retained in \emph{surrogate models} that are derived from the source model. 
Methods of derivation include modifying the source model, e.g., through fine-tuning~\cite{uchida2017embedding} or weight pruning~\cite{zhu2017prune}, and extraction of the source model, which uses a process related to knowledge distillation~\cite{hinton2015distilling} to train a different model. 

We differentiate between zero-bit and multi-bit watermarking.
Zero-bit watermarking encodes only the presence of a watermark, whereas multi-bit watermarking allows encoding a message containing several bits of information. 
For zero-bit watermarking schemes, we represent the message that can be extracted as a bit-string $m\subset\{0,1\}$, where $m_i=1$ means that the presence of the $i$-th watermark has been detected and $m_i=0$ otherwise. 
Note that the message embedded into the source model has zero bits because extracting from a source model after embedding should always return the message of all ones.
For multi-bit watermarking, the message $m\subset \{0,1 \}$ can be chosen by the user and thus contains multiple bits of information. 
A watermarking scheme can be formalized by an embedding and extraction procedure. 
\begin{itemize}
    \item \textbf{Embed}$(T, m, M)$: Takes a watermarking key $T$, a message $m\subset \{0,1\}$ and a model $M$ and outputs a marked model $\hat{M}$ embedded with a message $m$.
    \item \textbf{Extract}$(T, M)$: Takes a watermarking key $T$, a model $M$ and outputs the message $m\subset \{0,1\}$ extracted from model $M$ using key $T$. 
\end{itemize}
The watermarking key $T$ contains the secret information required to extract a watermark. 
For example, the watermarking key can consist of images~\cite{adi2018turning}, a bit-vector~\cite{uchida2017embedding} or a combination of both~\cite{rouhani2018deepsigns}. 
There exists a trivial procedure to verify whether a model $\hat{M}$ retains a watermark.
This verification procedure takes as parameters a watermarking key $T$ and message $m$, a model $\hat{M}$ and a decision threshold $\theta\in [0,1]$.
The decision threshold specifies the lowest tolerable success rate between message $m$ and the message extracted from $\hat{M}$ to verify whether the watermark is retained. 
The verification extracts a message $\hat{m}$ from model $\hat{M}$ using $T$ and computes the success rate $d=\Delta(\hat{m},m)$. 
If the watermark is retained ($d\geq \theta$) the verification outputs $b=1$ and $b=0$ otherwise.

\subsection{Watermarking Categories}
We systematize DNN watermarking schemes as a tree diagram in Figure~\ref{fig:tree_diagram}.
These schemes can be differentiated by (i) the watermark carrier, (ii) the stage at which the watermark is embedded, and (iii) whether the embedding requires access to a pre-trained source model for the generation of the watermarking key. 
The watermark carrier can be the model's parameters or its functionality. 
In the latter case, modification of the functionality can either occur during inference while the model is deployed or during training.  
If the embedding occurs during training, a watermarking scheme may require that the model is pre-trained.
In this case, the secret key's generation depends on the trained model, e.g., when the secret key contains adversarial examples~\cite{goodfellow2014explaining}.
Otherwise, the watermarking key can be generated independently of the model and only depends on the dataset.
In summary, for a systematic analysis of the robustness of watermarking, we propose the following four categories of watermarking schemes. 

\begin{enumerate}
% model dependent
    \item \textbf{Model Dependent}~\cite{le2020adversarial, chen2019blackmarks, jia2020entangled}:
    A model dependent scheme embeds the message into the model's functionality during training, where the watermark key samples depend on the model. 
    Watermarking schemes in this category either rely on adversarial examples~\cite{szegedy2013intriguing} generated for the source model~\cite{le2020adversarial, chen2019blackmarks} or use the source model to modify the watermarking key~\cite{jia2020entangled}. 
    
    \item \textbf{Model Independent}~\cite{adi2018turning, li2019piracy, zhang2018protecting}:
    A model independent scheme embeds the message into the functionality during training, where the watermarking key samples do not depend on the model. 
    The watermark is a \emph{backdoor}~\cite{gu2019badnets}, i.e., secret functionality learned by the source model from the training set. 
    A backdoor is embedded by injecting additional samples into the training set, and hence, the generation of the watermarking key does not depend on the source model. 
    \item \textbf{Active}~\cite{szyller2019dawn}: 
    An active scheme embeds the message into the model's functionality during inference. 
    It requires the defender to control the source model's deployment. 
    Active schemes only defend against attackers with black-box access to the source model by \emph{postprocessing} predictions returned by the source model on input queries. 

    \item \textbf{White-box} (Parameter Encoding)~\cite{chen2018deepmarks, rouhani2018deepsigns, uchida2017embedding}: % White-box, 
    A \emph{white-box} scheme embeds the message into the model's parameters~\cite{uchida2017embedding, chen2018deepmarks} or into the activations of its hidden layers~\cite{rouhani2018deepsigns}. 
    Verification requires white-box access to the source model, i.e., access to the model's parameters. 
\end{enumerate}

\subsection{Watermark Removal Attack Categories} 
A watermark removal attack takes as input the source model and outputs a surrogate model. 
It is successful if the surrogate model does not retain the watermark, and it has a similar utility (measured in test accuracy) as the source model. 
We survey (i) known removal attacks~\cite{uchida2017embedding, shafieinejad2019robustness, zhu2017prune, aiken2020neural, tramer2016stealing, wang2019neural, liu2018fine}, (ii) methods that derive a surrogate model but have not been evaluated as removal attacks against DNN watermarking~\cite{torrey2010transfer, hinton2015distilling, orekondy2019knockoff, madry2017towards, lin2019defensive, xu2017feature, xu2017feature, dziugaite2016study, zantedeschi2017efficient, lin2019invert, hubara2017quantized} and (iii) novel, adaptive attacks proposed in this paper.
We investigate which of these methods successfully remove watermarks. 
From all surveyed removal attacks, we derive the following three attack categories.
\begin{itemize}
    \item \textbf{Input Preprocessing}: 
    Input preprocessing attacks modify the data samples for classification before passing them through the surrogate model. 
    The attacker must have white-box access to the source model.
    \item \textbf{Model Modification}: Model modification attacks transform the source model's parameters, e.g., by fine-tuning~\cite{uchida2017embedding} or pruning~\cite{zhu2017prune}.
    The attacker must have white-box access to the source model. 
    \item \textbf{Model Extraction}: Model extraction attacks train a different surrogate model by transferring knowledge from the source model into the surrogate model. 
    The surveyed model extraction attacks need only black-box access to the source model, with the exception of knowledge distillation~\cite{hinton2015distilling} which requires white-box access. 
\end{itemize}

\subsection{Formalizing Watermarking Requirements }
\label{sec:form_watermark_req}
Ideal watermarking should satisfy the requirements listed in Table~\ref{tab:requirements}. 
We now formalize the two properties investigated in this paper: robustness and integrity. 
We refer to the watermark extraction procedure by $E(T, \hat{M})$ for ease of notation. 

\textbf{Robustness}: Robustness requires that a message extracted from a surrogate model is approximately the same as the message extracted from the source model. 
The following condition should hold for $\varepsilon \geq 0$, a model $M$, a watermarking key $T$, a message $m$ and any watermark removal attack $\mathcal{A}$. 
\begin{gather*}
    \hat{M} \leftarrow \text{Embed}(T, m, M) \\
    \Delta\big(E(T, \hat{M}), E(T, \mathcal{A}(\hat{M}))\big) \geq 1-\varepsilon
\end{gather*}
Note that robustness as defined is trivial by itself for zero-bit watermarking since the extraction algorithm could always return an all-ones message. 

\textbf{Integrity}: Integrity requires a low success rate between messages extracted from a marked model $\hat{M}$ and an unmarked model $M_0$. 
Given the watermarking key $T$, a message $m$, the marked model $\hat{M}$ as defined above and an unmarked model $M_0$ the following condition should hold for $\varepsilon \geq 0$.
\begin{align*}
    \Delta(E(T, \hat{M}), E(T, M_0)) \leq \varepsilon
\end{align*}
We evaluate whether DNN watermarking can satisfy robustness and integrity. 
In the next section, we define a generic adversary model and present all watermarking schemes and removal attacks evaluated in this paper. 

\begin{table*}[]
    \centering
        \begin{tabular}{|c|cccc|}
\hline
\textbf{Defense}                                                      & \textbf{Category}  & \textbf{Verification} &  \multicolumn{1}{c|}{\textbf{Capacity}}  \\ \hline
\multicolumn{1}{|c|}{\makecell{Adi~\cite{adi2018turning}}}                           & Model Independent  & Black-Box  & \multicolumn{1}{c|}{Multi-bit}          \\ \hline
\multicolumn{1}{|c|}{\makecell{Content~\cite{zhang2018protecting}, Noise~\cite{zhang2018protecting},\\ Unrelated~\cite{zhang2018protecting}}}   & Model Independent  & Black-Box  & \multicolumn{1}{c|}{Zero-bit}          \\ \hline
\multicolumn{1}{|c|}{Jia~\cite{jia2020entangled}, Frontier Stitching~\cite{le2020adversarial}}  & Model Dependent    & Black-box & \multicolumn{1}{c|}{Zero-bit}           \\ \hline
\multicolumn{1}{|c|}{Blackmarks~\cite{chen2019blackmarks}}          & Model Dependent    & Black-box  & \multicolumn{1}{c|}{Multi-bit}          \\ \hline
\multicolumn{1}{|c|}{\makecell{Uchida~\cite{uchida2017embedding}, Deepsigns~\cite{rouhani2018deepsigns}, \\DeepMarks~\cite{chen2018deepmarks}}}             & Parameter Encoding & White-box  & \multicolumn{1}{c|}{Multi-bit}          \\ \hline
\multicolumn{1}{|c|}{DAWN~\cite{szyller2019dawn}}                 & Active             & Black-box & \multicolumn{1}{c|}{Multi-bit}        \\\hline
\end{tabular}

    \caption{All watermarking schemes evaluated in this paper. See Appendix~\ref{sec:schemes} for a description of each method.\label{tab:watermarking_schemes}}
\end{table*}
\begin{table*}[]
    \centering
        \begin{tabular}{|c|ccccc|}
\hline
\textbf{Attack}    & \textbf{Category}  & \textbf{Deployment} & \multicolumn{1}{c|}{\textbf{Data}} \\ \hline
\multicolumn{1}{|c|}{\makecell{Input Reconstruction~\cite{lin2019invert}, JPEG Compression~\cite{dziugaite2016study}, Input Quantization~\cite{lin2019defensive},\\
Input Smoothing~\cite{xu2017feature}, Input Noising~\cite{zantedeschi2017efficient}, Input Flipping, Feature Squeezing~\cite{xu2017feature}}}  & \makecell{Input Preprocessing}  & \makecell{White-box}  & \multicolumn{1}{c|}{None}  \\\hline

% Model Modification
\multicolumn{1}{|c|}{\makecell{Adversarial Training~\cite{madry2017towards},  Fine-Tuning (RTLL, RTAL)~\cite{uchida2017embedding}, Weight Quantization~\cite{hubara2017quantized}, \\
 Label Smoothing~\cite{szegedy2016rethinking}, Fine Pruning~\cite{liu2018fine},  Feature Permutation (Ours), Weight Pruning~\cite{zhu2017prune}, \\
Weight Shifting (Ours), Neural Cleanse~\cite{wang2019neural}, Regularization~\cite{shafieinejad2019robustness}, Neural Laundering~\cite{aiken2020neural}}}        & \makecell{Model Modification} & \makecell{White-box}            & \multicolumn{1}{c|}{Domain}       \\ \hline

\multicolumn{1}{|c|}{Overwriting~\cite{uchida2017embedding}, Fine-Tuning (FTLL, FTAL)~\cite{uchida2017embedding}}     & \makecell{Model Modification} & \makecell{White-box}            & \multicolumn{1}{c|}{Labeled}           \\ \hline

% Input Preprocessing

% Model Extraction

\multicolumn{1}{|c|}{Knockoff Nets~\cite{orekondy2019knockoff}}           & \makecell{Model Extraction}            & \makecell{Black-box}            & \multicolumn{1}{c|}{Transfer} \\\hline
\multicolumn{1}{|c|}{Distillation~\cite{hinton2015distilling}}           & \makecell{Model Extraction}            & \makecell{White-box}            & \multicolumn{1}{c|}{Domain}   \\\hline
\multicolumn{1}{|c|}{\makecell{ Transfer Learning, Retraining~\cite{tramer2016stealing}, Smooth Retraining~(Ours)  \\ 
Cross-Architecture Retraining~(Ours), Adversarial Training (From Scratch)~\cite{madry2017towards}}}        & \makecell{Model Extraction} & \makecell{Black-box} & \multicolumn{1}{c|}{Domain}  \\\hline
\end{tabular}
    \caption{A list of all watermark removal attacks evaluated in this paper and the attacker's capabilities (see Section~\ref{sec:adversary_model}). 
    We refer to Appendix~\ref{sec:attacks} for a more detailed description of the attacks and their parameters used for our ablation study. RTAL and RTLL use predicted labels, whereas FTAL and FTLL use ground-truth labels (otherwise, gradients are zero).  \label{tab:removal_attacks}
    }
\end{table*}

\section{Adversary Model}
\label{sec:adversary_model}
In this section, we describe the attacker's goals and capabilities. 
Our study covers many different watermarking schemes and removal attacks that assume different adversary models. 
For example, model modification attacks require white-box access to the source model, whereas many model extraction attacks only require black-box access. 
We present a generic adversary model for any watermarking scheme and watermark removal attack. \Cref{tab:watermarking_schemes,tab:removal_attacks} summarize the defender's and attacker's capabilities for all methods surveyed in this paper.

\subsection{Attacker's Goals}
\label{sec:attackersgoals}
The attacker's primary goal is to derive a surrogate model from access to the source model (i) without the retained watermark that is (ii) \emph{well-trained}, i.e., it has a similar test accuracy as the source model.
A secondary goal is to reduce resources needed for the removal attack, such as the attack's computation time. 
We formalize a security game between the attacker and the defender. 
Given a secret watermarking key $T$ and message $m$, only known to the defender, two well-trained, unmarked models $M, M_0$ and a watermark removal attack $\mathcal{A}$, the security game can be formalized as follows for $\varepsilon \geq 0$. 
\begin{enumerate}
    \item Train $M$ and $M_0$ and send $M$ to the defender.
    \item Defender embeds the watermark $\hat{M} \leftarrow \text{Embed}(T, m, M)$
    \item Attacker derives the surrogate model $M_1 \leftarrow \mathcal{A}(\hat{M})$
    \item Sample $M_b \xleftarrow[]{\$} \{M_0,M_1\}$ and send $M_b$ to the defender
    \item Attacker wins if:
        \begin{align*}
            \text{Pr}[\text{Verify\footnotemark}(T, M_b)=b] \leq 0.5 + \varepsilon
        \end{align*}
        \footnotetext{The process 'Verify' checks if the success rate of the embedded and extracted message is higher than the decision threshold (see Section~\ref{sec:defining_watermarking})}
\end{enumerate}
The robustness and integrity of a watermarking scheme are violated if an attacker can win this security game.

\subsection{Attacker's Capabilities. }
\label{sec:attacker_capabilities}
We now present the capabilities of an attacker in the form of a unified adversary model. 
Tables~\ref{tab:watermarking_schemes} and \ref{tab:removal_attacks} summarize the adversary model for each watermarking scheme and removal attack surveyed in this paper. 

\textbf{Deployment. }
The deployment property summarizes the access of the attacker to the source model's parameters.
It is white-box if all of the source model's parameters are accessible to the attacker and black-box if only the source model's API is accessible. 
Note that an attacker with white-box access is more informed and can also invoke attacks of an attacker who only has black-box access. 

\textbf{Dataset. } 
The dataset property summarizes the availability of an auxiliary dataset to the attacker.
Many attacks from related work require at least the availability of unlabeled domain data, and some even need access to data where a subset is labeled with ground-truth labels.
We assume the attacker is limited in the amount of labeled data; otherwise, they could train their own model and would not need to steal the defender's source model.
From all attacks, we identify the availability of the following three datasets to the attacker. 
\begin{enumerate}
    \item \textbf{Labeled}: Data from the same distribution where a subset of at most a third of the data is labeled.
    \item \textbf{Domain}: Unlabeled data from the same distribution.
    \item \textbf{Transfer}: Labeled data from a different distribution.
\end{enumerate} 
An attacker with access to a subset of labeled data is more informed than an attacker with access to only domain data. 
We consider collecting labeled data from a different distribution, and in all of our experiments, we use the Open Images~\cite{openimages} dataset as our transfer set. 

\textbf{Speed. } Throughout the paper, we assume unbounded computational resources for the attacker.
We only measure the runtime of attacks for a discussion of the practicality of the attack.
Attacks are categorized concerning the total training time of an unmarked model from scratch. 
We consider an attack to be \emph{fast} if it requires less than $25\%$ of the training time, \emph{medium} for times between $25\%$ and $75\%$ and \emph{slow} for longer runtimes. 
We categorize speed according to the attack's runtime on the highest resolution dataset investigated in this paper (i.e., ImageNet~\cite{imagenet}).

\section{Watermarking Schemes}
This section contains survey-style descriptions of the investigated watermarking schemes. 
We refer to a watermarking scheme by the first author's name for simplicity unless it is known under a different name.

\subsection{Model Independent}

\textbf{Adi}~\cite{adi2018turning} uses a secret watermarking key consisting of abstract, out-of-distribution images. 
A label for an image is randomly sampled over all classes, excluding the image's true label. 
The embedding consists of fine-tuning the model on the watermarking key.
A message is extracted from a surrogate model by requesting labels for the watermarking key images. 

\textbf{Zhang}~\cite{zhang2018protecting} proposes three different schemes, referred to as \emph{Content}, \emph{Noise} and \emph{Unrelated}.  
These three schemes differ only in their selection of the watermarking keys.
The watermarking keys are selected as follows. 
\begin{itemize}
    \item \textbf{Content}:
    The secret watermarking key are images sampled from one \emph{source} class that are perturbed by a secret, additive mask (e.g., a white square covering part of the image). 
    The same mask is used for all watermarking keys. 
    \item \textbf{Noise}:
    The scheme uses a secret, additive mask that consists of random, Gaussian noise. 
    \item \textbf{Unrelated}:
    The watermarking key are images sampled from a different domain that is unrelated to the source model's domain.
    For example, if a model is trained to classify animal species, the watermarking key could contain images of automobiles. 
\end{itemize}
All watermarking key images are labeled with the same \emph{target} class that is sampled randomly. 
The embedding and extraction is the same as Adi for all three schemes.  

\subsection{Model Dependent}

\textbf{Frontier-Stitching}~\cite{le2020adversarial} uses a watermarking key consisting of adversarial examples~\cite{szegedy2013intriguing}.
Adversarial examples are images that have been modified (often imperceptibly) to trigger a misclassification when a DNN predicts the image's label. 
The authors generate these adversarial examples using the Fast Gradient Method (FGM)~\cite{goodfellow2014explaining} and the pre-trained source model. 
This method has a given probability of failure, meaning that its output is not adversarial and is correctly classified by the DNN. 
The watermarking key is composed of such \emph{false} adversarial examples and equally many \emph{true} adversarial examples. 
All adversarial examples are labeled by the ground-truth label in the watermarking key. 
The embedding and extraction process is the same as Adi. 

\textbf{Blackmarks}~\cite{chen2019blackmarks} also relies on adversarial examples, similar to Frontier-Stitching. 
The authors propose a pre-processing step that clusters all class labels into two groups using k-means clustering on the pre-trained source model's logit activations. 
These clusters will be used to encode bits. 
The idea is to randomly select images from one cluster and use a targeted adversarial attack so that the source model predicts any class from the other cluster. 
During embedding, an additional loss term is introduced that minimizes the bit error rate between the predicted cluster and the assigned cluster of the trigger. 
The authors also present a method to mitigate unintended \emph{transferability} of the watermarking key images. 
An adversarial example is transferable if it is adversarial to many models, i.e., it is not only adversarial to the source model for which it has been generated. 
The embedding and extraction is similar to Adi, except that a new loss term is added during embedding as described above. 

\textbf{Jia}~\cite{jia2020entangled} proposes using the soft nearest neighbor loss (SNNL)~\cite{snnl1, snnl2} as an additional loss during training to \emph{entangle} feature representations of the watermark with the training data. 
Two groups are entangled if the average distance between their elements is lower than the average distance within each group, which the SNNL measures. 
A temperature parameter controls the significance of short and long distances for the total loss and can be tuned during training. 

The watermark generation defines two groups of elements belonging to a source and a target class.
These classes are chosen by computing two classes with the highest similarity using the pre-trained source model's hidden activations. 
All elements from a source class are modified by adding a (secret) trigger pattern and changing their label to the target class, similar to the Content watermark described earlier. 
The trigger is added at the location where the gradient (backpropagated through the source model) with respect to the SNNL is highest. 

The embedding uses alternating training on batches of watermarking images and images from the primary task. 
Additionally, the author's propose automatically updating the temperature parameter during training. 

\subsection{Active}

\textbf{DAWN}~\cite{szyller2019dawn} proposes a scheme where a low proportion $r$ of all predictions are randomly relabeled and added to the watermarking key.
The authors implement a method that recognizes similar samples (as perceived by the source model) and returns the same label when the attacker tries to query a sample twice. 
This similarity detection is implemented by using the activation of some target layer of the source model. 
The watermarking key consists of images from the attacker's dataset, and labels are assigned randomly. 

\subsection{Parameter Encoding}

\textbf{Uchida}~\cite{uchida2017embedding} propose embedding a message into the weights of some \emph{target} convolutional layer. 
The idea is to add an \emph{embedding} loss during training that regularizes the model and is minimized when the message can be extracted successfully and with a large margin. 
Let $W \in \mathbb{R}^{n\times c \times w\times}$ be the convolutional filters of a target layer, where $n$ is the number of filters, $c$ are the number of channels, and $w,h$ are the width and height of each filter. 
The scheme computes a mean filter $\bar{W} = \frac{1}{n} \sum_{i=1..n} W_i$ to deal with the permutation invariance of the filters. 
This mean filter is flattened $\hat{W}\in \mathbb{R}^{(c \cdot w \cdot h)}$ and a random projection matrix $A \in \mathbb{R}^{k \times (c\cdot w \cdot h)}$ is sampled, where $k$ is the desired key length. 
The embedding consists of fine-tuning using the embedding loss described above. 
A message can be extracted by computing $m'=A\hat{W}^T$ and applying the following rule. 
\begin{align}
    m_i = \begin{cases}
1 & m'_i \geq 0 \\
0 &\text{otherwise}
\end{cases}
\end{align}

\textbf{DeepMarks}~\cite{chen2018deepmarks}\footnote{DeepMarks is labeled as a fingerprint by the authors, but since it modifies the model by embedding a message, it is a watermark as per our definition.} proposes a similar embedding method as in Uchida, with a notable difference in how a message is extracted. 
The authors compute the dot product between $A\hat{W}$ and the owner's signature, which returns a correlation score between $-1.0$ and $1.0$. 
A correlation of $1.0$ is interpreted as a perfectly retained watermark, whereas a correlation $\leq 0$ corresponds to a watermark accuracy of zero. 

\textbf{DeepSigns}~\cite{rouhani2018deepsigns} proposes a watermarking scheme that uses the activations of some target layer of the source model to encode a message.
The intuition is that features extracted from samples belonging to the same \emph{source} class form clusters, whose properties can be used to embed watermarking information.
In the author's paper, clusters are modelled using a Gaussian Mixture Model, whereby each feature cluster $c_i$ is described by a mean $\mu_i$ and a standard deviation $\sigma_i$. 
An embedding loss is added during training that modifies each cluster's mean and allows embedding $n$ bits of information per cluster, i.e., for $m$ clusters, we can embed a message with $m\cdot n$ bits. 
A random projection matrix is sampled that is multiplied with the source model's activation on the target layer given the watermarking images. 
The result is a binary vector that is then activated by the Sigmoid function. 
Elements of this binary vector can be interpreted as binary digits by comparing their values with a threshold, similar to Uchida's extraction.  

\section{Removal Attacks}
This section provides summaries of all removal attacks surveyed in this paper. 
The list contains (i) known watermark removal attacks, (ii) attacks that have not been evaluated as removal attacks and (iii) novel removal attacks. 

\subsection{Input Preprocessing}

\textbf{Input Reconstruction}~\cite{lin2019invert} uses an autoencoder\footnote{\href{https://github.com/foamliu/Autoencoder}{https://github.com/foamliu/Autoencoder}}~\cite{ng2011sparse} to compress and reconstruct images before passing them to the surrogate model.
The idea is that an autoencoder trained on non-watermarked images will fail at reconstructing artifacts in the image it has never seen before (such as additive masks or random noise) while preserving the remaining image's content. 

\textbf{Input Noising}~\cite{zantedeschi2017efficient} adds Gaussian noise with zero mean and some standard deviation to the entire image.

\textbf{Input Quantization}~\cite{lin2019defensive} quantizes the pixel values for an input images. 
For a given number of bits $b$, the input space is quantized into $2^b$ evenly spaced intervals. 
Every pixel of the input image is set to the mean of its interval. 

\textbf{Input Smoothing}~\cite{xu2017feature} convolves some kernel over the image, which makes the image appear more blurry. 
Possible kernels are a mean, median, and Gaussian kernel. 
The image appears more blurry after applying the convolution. 

\textbf{Input Flipping} flips an image along its horizontal axis. 

\textbf{JPEG Compression}~\cite{dziugaite2016study} is similar to input reconstruction, but instead of using an autoencoder, the image is compressed through the JPEG compression algorithm. 

\textbf{Feature Squeezing}~\cite{xu2017feature} is similar to input quantization, except that input values are rounded to the nearest quanta, and the quanta values are chosen to be multiples of $0.5^k$ for some $k\in \mathbb{N}$.

\subsection{Model Modification}
\textbf{Adversarial Training}~\cite{madry2017towards} is a method to increase a model's robustness to adversarial examples.
A random subset of the training dataset is perturbed using the Projected Gradient Descent (PGD)~\cite{madry2017towards} adversarial attack. 
These examples are then injected into the training dataset and the model is fine-tuned on these examples with their ground-truth labels. 

\textbf{Feature Permutation}. DNNs are invariant to feature permutations, meaning that neurons in a hidden layer can be permuted without affecting the model's functionality. 
We use (random) feature permutation as an adaptive attack designed specifically against Deepsigns~\cite{rouhani2018deepsigns}, which encodes the message into the activations of hidden layers. 

\textbf{Fine-Pruning}~\cite{liu2018fine} is designed for \emph{backdoor removal} that first prunes dormant neurons and then fine-tunes the model to regain the drop in test accuracy.
The idea is that neurons which are never highly activated for benign inputs likely implement the functionality of the backdoor. 
Such neurons are eliminated by setting their activation to zero. 

\textbf{Fine-Tuning}~\cite{uchida2017embedding} as a model stealing attack refers to a set of attacks that first apply a transformation to the model, followed by fine-tuning. 
\begin{itemize}
    \item Fine-Tune All Layers (\textbf{FTAL}). All weights are fine-tuned. 
    \item Fine-Tune Last Layer (\textbf{FTLL}). All but the last layer's weights are frozen while the model is fine-tuned.
    \item Retrain All Layers (\textbf{RTAL}). The last layer's weights are re-initialized, and all weights are fine-tuned. 
    \item Retrain Last Layer (\textbf{RTLL}). The last layer's weights are re-initialized, and only that layer's weights are fine-tuned. 
\end{itemize}
RTAL and RTLL use predicted labels, whereas FTAL and FTLL use ground-truth labels (otherwise, gradients are zero). 

\textbf{Overwriting}~\cite{uchida2017embedding} embeds a watermark using the same watermarking scheme but a different watermarking key.

\textbf{Label Smoothing}~\cite{szegedy2016rethinking} is a regularization method that computes the weighted mean of a uniform distribution over all labels with a one-hot or (in our case) predicted label by the source model. 
The idea is to regularize the model by making classes other than the ground-truth class appear likely. 

\textbf{Regularization}~\cite{shafieinejad2019robustness} is a two-phased attack that strongly regularizes the model in the first phase, leading to a drop in test accuracy, which is compensated by fine-tuning the model in the second phase.
The idea of the regularization phase is to move the model's parameter far away from their origin, while the fine-tuning phase finds a (different) local minima. 

\textbf{Neural Cleanse}~\cite{wang2019neural} is an attack designed for \emph{backdoor removal}. 
It first reverse-engineers the watermark trigger and then removes ('unlearns') the trigger from the model.
When a trigger has been reverse-engineered, (i) it can be unlearned through fine-tuning on different labels or (ii) removed by pruning most activated neurons in some layer. 

\textbf{Neural Laundering}~\cite{aiken2020neural} is an extension of Neural Cleanse, specifically designed against model independent watermarking schemes. 
The authors re-use the reverse-engineering trigger step from Neural Cleanse
and then iteratively prune weights triggered by the backdoor.

\textbf{Weight Pruning}~\cite{zhu2017prune} randomly prunes the weights in the source model until a given sparsity $\rho$ is reached in each layer. 

\textbf{Weight Shifting} is a novel, adapted attack against white-box, parameter encoding watermarking schemes. 
The idea is to apply a small perturbation to all filters of each convolutional layer in the network, followed by fine-tuning the model to regain the loss in test accuracy. 
We design weight shifting as an efficient and effective model stealing attack specifically against Uchida~\cite{uchida2017embedding} and Deepmarks~\cite{chen2018deepmarks}. 

We explain the attack's idea at the example of Uchida, but a similar intuition holds for Deepmarks where the extraction is highly similar. 
A weakness of Uchida exploited by weight shifting is that the attacker knows that if all filters were inverted, i.e. $W_i' = -W_i$, then the watermark accuracy would be zero. 
We cannot directly invert all filters, as the model experiences a significant drop in test accuracy.
Hence, we construct a 'softer' version of the attack that only moves each filter in the direction of the inverse mean multiplied by some constant weight parameter $\lambda_1\in \mathbb{R}$. 
We additionally add small random Gaussian noise to each filter to encourage the network to find slightly different filters in the fine-tuning phase. 

Our attack can be formalized by the function $S(W; \lambda_1, \lambda_2)$, which takes as input a set of filters $W$ and outputs a shifted set of filters $W'$. 
The parameter $\lambda_1, \lambda_2$ trade off the attack's efficiency with its effectiveness. 
Let $A$ be a random normal matrix of the same shape as each filter $W_i$ with a variance equivalent to the variance over all filters for a convolutional layer and a mean of zero.  
Shifted weights for each convolutional layer can be computed by applying the following function. 
\begin{align}
    \text{S}(W; \lambda)_i = W_i - \frac{\lambda_1}{n} \sum_{j=1..n} W_j - \lambda_2 A
\end{align}

\textbf{Weight Quantization}~\cite{hubara2017quantized} compresses a model by quantizing its weights. 
This attack is equivalent to Input Quantization, but quantizes the model's weights rather than the input image. 

\subsection{Model Extraction}
\textbf{Retraining}~\cite{tramer2016stealing} is a method to train a surrogate model from API access to a source model.
The surrogate model is trained from scratch on the softmax output of the source model.
Cross-Architecture retraining is equivalent to retraining, but the surrogate and source model's architecture differs. 

\textbf{Cross-Architecture Retraining} is equivalent to retraining, but the surrogate model uses a different DNN architecture than the source model. 
The idea is to increase the dissimilarity to the source model, which intuitively decreases the likelihood that the watermark is transferred to the surrogate model. 

\textbf{Distillation}~\cite{hinton2015distilling} is a method originally designed as a model compression technique that distills knowledge from a (larger) teacher DNN to a (smaller) student DNN. 
The idea is to scale the teacher DNN's logit activations before applying the softmax function, which increases the prediction's entropy. 
The student is trained using this scaled prediction vector. 

\textbf{Smooth Retraining} retrains a surrogate model on smoothed labels obtained from querying the source model for multiple variations of the same image. 
For each query, a random, affine transformation (e.g., random cropping) is applied to the image, and the mean of all received labels is computed as the final label.
We design smooth retraining as an adaptive attack against the active watermarking scheme DAWN.
The intuition is that if DAWN responds with a false label for one image, variations of the same image have a high probability of receiving the label predicted by the source model. 
     
\textbf{Knockoff Nets}~\cite{orekondy2019knockoff} is a method for training a surrogate model from scratch using only data from a different domain, referred to as the \emph{transfer set}. 
We implement the random selection approach proposed by the authors. 

\textbf{Transfer Learning}~\cite{torrey2010transfer} is similar to Retraining, but instead of using a randomly initialized model, the attack assumes access to a pre-trained surrogate model from a different domain. 
This surrogate is fine-tuned on data from the source model's domain using labels predicted by the source model.
The pre-trained surrogate model has already learned semantically meaningful filters for a different task. 
A common technique is to freeze the surrogate model's weights that are located in the lower layers (i.e., not updating them during training) to reduce the training time. 
Another option is to reduce the learning rates for lower layers to encourage re-use of these filters. 

\textbf{Adversarial Training (from scratch)}~\cite{madry2017towards} is equivalent to adversarial training described earlier, except that the attacker trains the surrogate model from scratch. 

\section{Measured Quantities}
\label{sec:methodology}
In this section, we present the measured quantities for conducting our experiments and describe the criteria for a watermark to be considered robust. 
Quantities, such as the test accuracy or an attack's runtime, are measured for the outcome of each removal attack against every watermarking scheme. 
We describe a method to empirically determine a decision threshold~(see Section~\ref{sec:background}) for each watermarking scheme and dataset. 
We introduce the \emph{Nash equilibrium} as a method to determine the best choice of parameters in an adversarial setting. 
The Nash equilibrium is computed over multiple parameter configurations for each scheme and removal attack.
Our goal is to empirically determine whether watermarking schemes are robust to removal attacks. 

\subsection{Measurements}
\label{sec:measurements}
First, we describe the quantities measured for each experiment and our processing of these measurements to ensure comparability between watermarking schemes.

\textbf{Embedding and Stealing Losses}. 
We measure the \emph{embedding} and \emph{stealing losses} as differences in test accuracy between an unmarked and a marked model and between a marked and a stolen surrogate model. 
The test accuracy is the accuracy of a model's predictions on an unseen, labeled dataset from the same distribution. 
First, we define an auxiliary function that computes the accuracy of a model $M$ on a dataset $D\subseteq\mathcal{X}\times \mathcal{Y}$.
\begin{align*}
    \text{acc}(M, D) &= \underset{(x, y) \in D}{Pr} [\underset{i}{\argmax}(M(x)) = \underset{j}{\argmax}(y)]
\end{align*}
The embedding loss is the difference in test accuracy between an unmarked model $M_0$ and a marked source model $\hat{M}$ on a labeled test dataset $D_{val} \subseteq \mathcal{X} \times \mathcal{Y}$.
\begin{align*}
    L_{embed}(M_0, \hat{M}, D_{val}) = \text{acc}(M_0, D_{val}) - \text{acc}(\hat{M}, D_{val}) 
\end{align*}
The stealing loss is the difference in test accuracy between a marked source model $\hat{M}$ and a stolen surrogate model $M_S$.
\begin{align*}
    L_{steal}(\hat{M}, M_S, D_{val}) = \text{acc}(\hat{M}, D_{val}) - \text{acc}(M_S, D_{val}) 
\end{align*}
The defender wants to minimize the embedding loss and the attacker wants to minimize the stealing loss. 

\textbf{Watermark Accuracy}.
The watermark accuracy is equal to the success rate defined in Section~\ref{sec:defining_watermarking}. 
We define the watermark accuracy for a surrogate model $\hat{M}$ and the message $m$ embedded into the source model using the secret watermarking key $T$.
Let $E$ be the message extraction function described in Section~\ref{sec:form_watermark_req}.
\begin{align*}
    \text{wmacc}(\hat{M}, m) = \Delta(E(T, \hat{M}), m)
\end{align*}
\textbf{Decision Threshold.}
The decision threshold $\theta \in [0, 1]$ determines the lowest tolerated watermark accuracy to verify that a watermark is retained in a model. 
Ideally, a scheme defines a decision threshold as part of their adversary model that we could use to assess its robustness.
Unfortunately, such methods are missing from the surveyed papers, meaning that we have to find a methodology to empirically derive decision thresholds for each watermarking scheme. 

Determining the decision threshold for a watermarking scheme is difficult.
The decision threshold depends on the watermark accuracy of an unmarked model, which can be influenced by factors such as the model's architecture or the randomness during training. 
For example, consider the case of the zero-bit, model dependent watermarking scheme Frontier Stitching~\cite{le2020adversarial}.
The presence of a watermark is detected if a surrogate model predicts the ground-truth labels for images that are part of the watermarking key. 
The watermarking key is composed of adversarial examples~\cite{szegedy2013intriguing} generated for the source model.
During the embedding, the source model is adversarially trained~\cite{madry2017towards} to predict ground-truth labels for the watermarking key, whereas unmarked models still likely predict incorrect labels if the adversarial examples are transferable~\cite{tramer2017space}.
The problem is that the watermark accuracy of an unmarked model can increase without access to the source model by using adversarial training. 
This affects this watermarking scheme's decision threshold, which should be chosen large enough so that unmarked models are not incorrectly verified. 
The challenge lies in estimating the cumulative probability distribution that an unmarked model has a watermarking accuracy larger than some decision threshold. 
Such an estimation enables determining a decision threshold so that an incorrect verification (i.e., falsely claiming that a watermark is retained in a model) has a given probability.

\textbf{Modeling the Decision Threshold.}
We empirically estimate an unmarked model's watermark accuracy given two random variables: the unmarked model and the watermarking key. 
Our goal is to estimate the cumulative probability that the watermark accuracy of a randomly generated watermarking key and a randomly sampled unmarked model is higher than some threshold. 
We make an i.i.d. assumption for our random variables and randomly generate 100 watermarking keys, each with a bit-length of $N=100$.
Then, we compute the watermark accuracy on a set of $30$ unmarked models for CIFAR-10 and $20$ unmarked models for ImageNet for every key and model pair. 
We model the cumulative normal probability distribution for the expected number of matched bits and choose a decision threshold. 
For our experiments, we choose a p-value of 0.05.
Table~\ref{tab:decision_threshold} shows a summary of the resulting decision thresholds for CIFAR-10 and ImageNet. 
We observe that some decision thresholds are different between CIFAR-10 and ImageNet, which requires the defender to derive a threshold specific to the model and dataset they want to protect. 
For the watermarking schemes Content, Noise, Frontier Stitching and Blackmarks, we observed that the choice of parameters affects their decision thresholds. 
In these cases, Table~\ref{tab:decision_threshold} shows the largest computed decision threshold, and we refer to Appendix~\ref{sec:appendix_decision_threshold} for more information.

\begin{table*}[]
    \centering
    \begin{tabular}{|c|c|c|c|c|c|c|c|c|c|c|c|}
\hline
                  & \textbf{Content} & \textbf{Noise} & \textbf{Unrelated} & \textbf{Adi} & \textbf{Jia} & \textbf{FS} & \textbf{Blackmarks} & \textbf{Deepmarks} & \textbf{Deesigns} & \textbf{Uchida} & \textbf{Dawn} \\ \hline
\textbf{CIFAR-10} & 0.0717             & 0.4867          & 0.1485              & 0.1504        & 0.0518         & 0.5330                        & 0.6225               & 0.3964              & 0.5254             & 0.5798           & 0.1641         \\ \hline
\textbf{ImageNet} & 0.0018             & 0.0229           & 0.0074               & 0.0066         & 0.1638        & 0.7164                       & 0.8073                & 0.3183              & 0.5848             & 0.5817          & 0.0061          \\ \hline
\end{tabular}
    \caption{This table shows the empirically determined, unscaled decision thresholds for each watermarking scheme on two datasets with a p-value of $0.05$. 
    We obtain these decision thresholds by generating 100 watermarking keys with a key length of $N=100$ each and compute the mean watermark accuracy on a set of unmarked models. 
    We use $30$ unmarked models for CIFAR-10 and $20$ models for ImageNet.
    We refer to Appendix~\ref{sec:appendix_decision_threshold} for details on the computation of the decision thresholds. 
    \label{tab:decision_threshold}}
\end{table*}

\textbf{Rescaling Watermark Accuracies.}
Our goal is to compare the robustness of different watermarking schemes.
Relating watermark accuracies from different schemes with each other is difficult because their decision threshold may differ.
In such cases, the watermark accuracy alone does not indicate whether a scheme is robust without knowledge of the scheme's decision threshold. 
We avoid this issue by linearly rescaling the watermark accuracy by the scheme's decision threshold $\theta$ so that a watermark is retained if the \emph{rescaled} watermark accuracy is at least equal to some fixed value $\theta'=0.5$ and removed otherwise.
This allows us to plot the watermark accuracies for different schemes into the same graph.  
We define a linear scaling function $S(x;\theta)$ that rescales the watermark accuracy so that (i) $S(\theta;\theta)=\theta'$ and (ii) $S(1;\theta)=1$. 
The rescaling function uses the scheme's (unscaled) decision threshold $\theta$ as a parameter and returns the scaled watermark accuracy. 
\begin{align}
    \label{formula:rescaling}
    S(x;\theta) = \max(0, \frac{1-\theta'}{1-\theta}x + \frac{\theta'-\theta}{1-\theta})
\end{align}
We clip the output to avoid negative watermark accuracies. 
From this point forward, unless stated otherwise, we only refer to the rescaled watermark accuracy and decision threshold. 

\textbf{Runtime}.
The runtime helps assess the practicality of a watermarking scheme or removal attack. 
We measure the runtime to (i) embed the watermark and (ii) run a removal attack.
Since runtimes depend on the hardware, we report all runtimes measured on (single) Tesla P100 GPUs. 

\textbf{Attack Success Criterion.} 
A success criterion determines whether a removal attack was successful in removing a watermark. 
We consider the watermark accuracy and the stealing loss of the surrogate model.
We say a removal attack was successful when the surrogate model's watermark accuracy is lower than the scheme's decision threshold and the surrogate model is well-trained. 
In our paper, we consider a maximum stealing loss of \emph{five} percentage points for a surrogate model to be considered well-trained. 
We refer to Section~\ref{sec:attackersgoals} for a security game that formalizes our success criterion. 

\subsection{Nash Equilibrium}
\label{sec:nash_equilibrium}
Our empirical analysis performs an ablation study over multiple sets of parameters for each watermarking scheme and removal attack. 
We now describe a method to measure the robustness of a watermarking scheme against one or more removal attacks under the consideration that the defender and attacker can choose from a set of parameters. 
For every watermarking scheme and removal attack, we ablate over multiple parameters (see Appendix~\ref{sec:schemes} and \ref{sec:attacks}) from which the defender and attacker can choose. 
We define a zero-sum game between the defender and attacker, where both players want to choose optimal parameters to maximize their gains.

We construct a \emph{payoff} matrix $V\in\mathbb{R}^{m\times n}$ for $n$ watermarking scheme parameters $\{d_0,..,d_n\}$ and $m$ removal attack parameters $\{a_0,..,a_m\}$.
The defender and attacker have full knowledge of this payoff matrix.
An entry in this matrix is computed by applying a payoff function on the outcome of running an attack with the row's parameters against a watermarking scheme with the column's parameters. 
We define the following payoff function. 
The payoff is zero for non-successful attacks, and otherwise, the payoff is equal to the surrogate model's test accuracy. 
At the start of the game, both players choose their strategy from the payoff matrix.
We observe that the defender maximizes their gain if they minimize the payoff, whereas the attacker wants to maximize the payoff. 
A \emph{Nash equilibrium} is found when neither player gains from changing their chosen parameters. 
Optimal parameters for both players can be derived as follows.
\begin{align}
    (d^*,a^*) = (d_i, a_j) = \argmin_{i} (\argmax_{j} V[i, j])
\end{align}
Using the Nash equilibrium to present our results, we demonstrate that successful watermark removal attacks exist due to the watermarking scheme's vulnerability rather than a wrong choice of parameters.  

\section{Experiments}
\label{sec:experiments}
In this section, we present the results of our experiments. 
We describe our experimental setup, a methodology for splitting data between the attacker and defender, and the model architectures. 
Then, we report measured quantities of the attacks and schemes, such as their runtimes or the embedding loss. 

We analyze the robustness of each watermarking scheme against (i) all attacks, (ii) categories of attacks, and (iii) individual attacks. 
The first experiment validates whether a scheme is robust if the attacker knows which scheme the defender has chosen (but not its parameters).
The second experiment analyzes which attack categories are most effective against each watermarking scheme.
The third experiment focuses on finding \emph{dominant} attacks, i.e., successful removal attacks that remove any watermark. 
Our results show that none of the single attacks on their own removes all watermarks. 
Still, we can find \emph{combined} attacks that are dominant.
We cannot depict all evaluation results in this paper.
Hence, we will make our results publicly available via an interactive graph that shows the Nash equilibrium for a set of attacks against a set of watermarking schemes\footnote{\url{https://crysp.uwaterloo.ca/research/mlsec/wrt}}.

\begin{figure*}
    \centering
	\subfloat[]{\raisebox{-0.5\height}{\includegraphics[width=.33\linewidth]{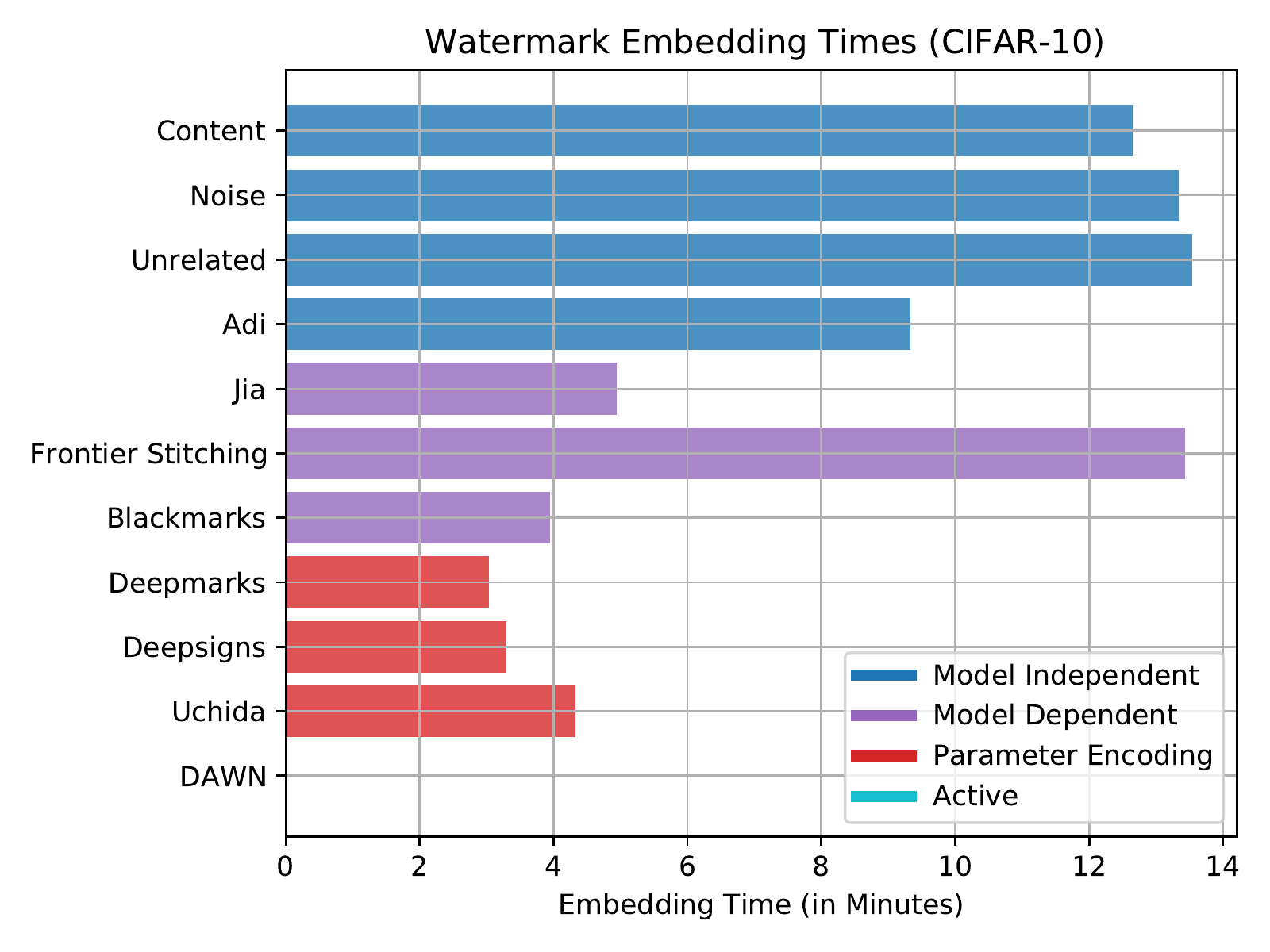}}\label{fig:cifar_embedding_time}}
	\subfloat[]{\raisebox{-0.5\height}{\includegraphics[width=.33\linewidth]{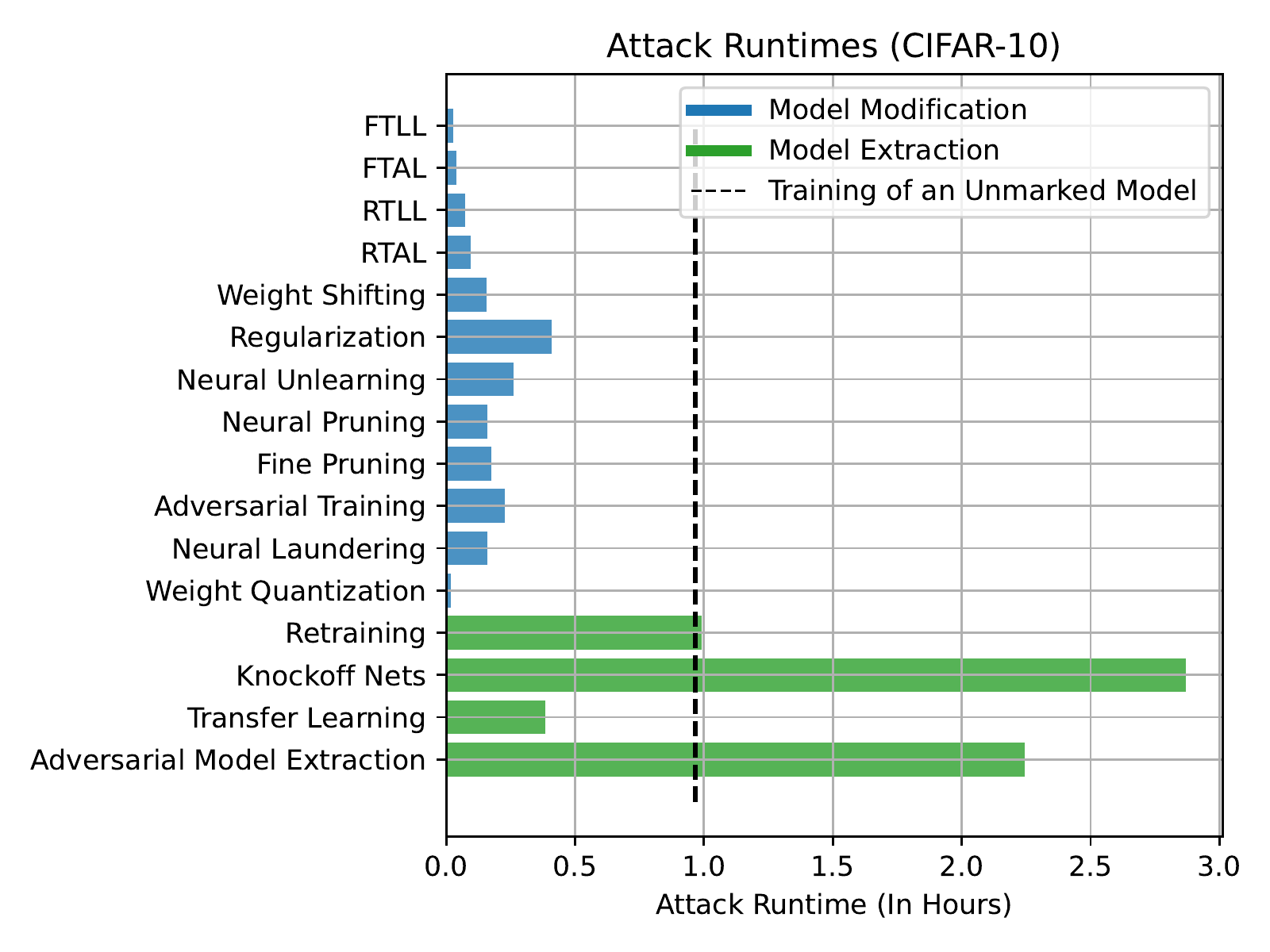}}\label{fig:cifar_attack_time}}
	\subfloat[]{\raisebox{-0.5\height}{\includegraphics[width=.33\linewidth]{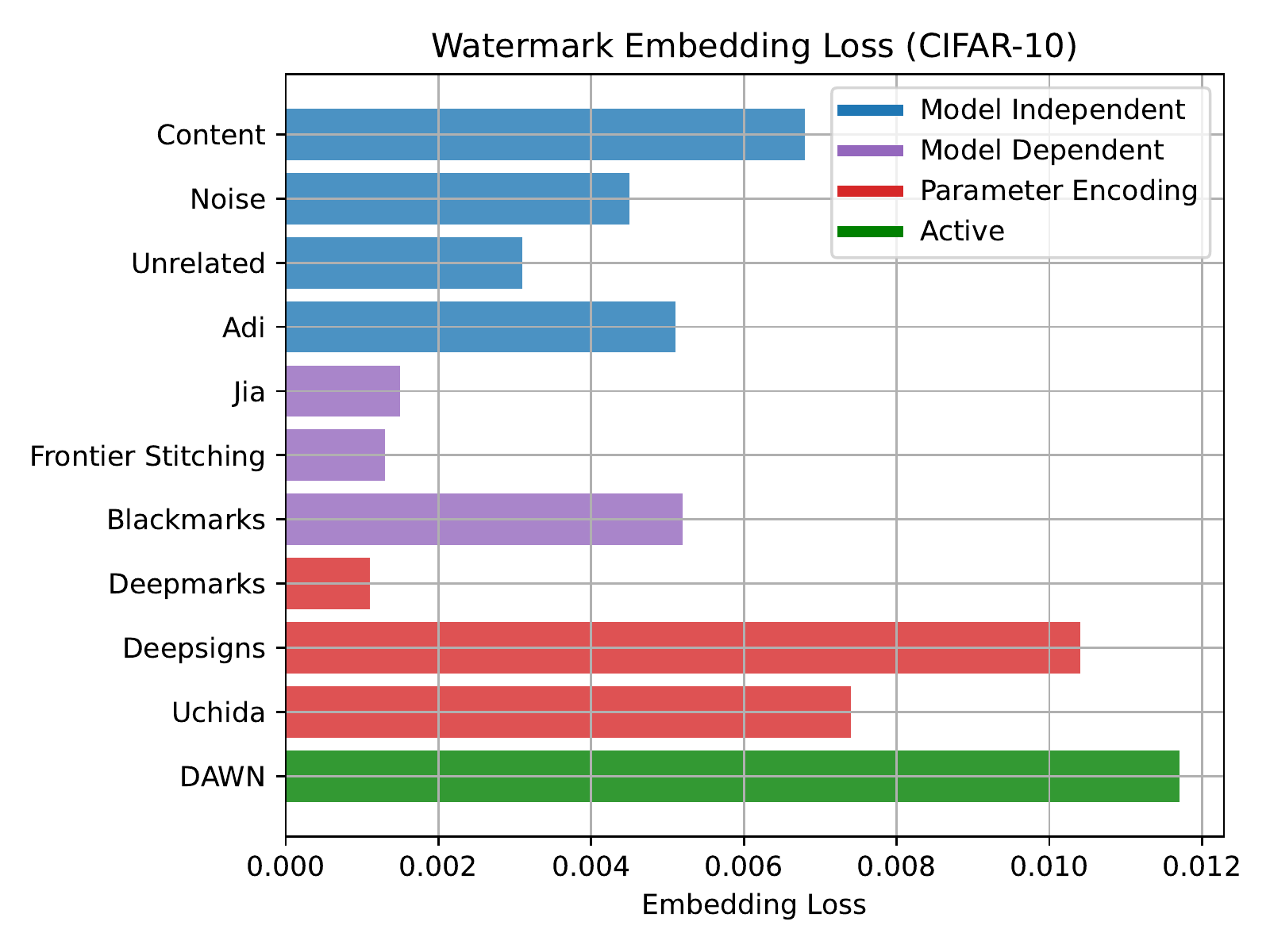}}\label{fig:cifar_embedding_loss}}\\
	\subfloat[]{\raisebox{-0.5\height}{\includegraphics[width=.33\linewidth]{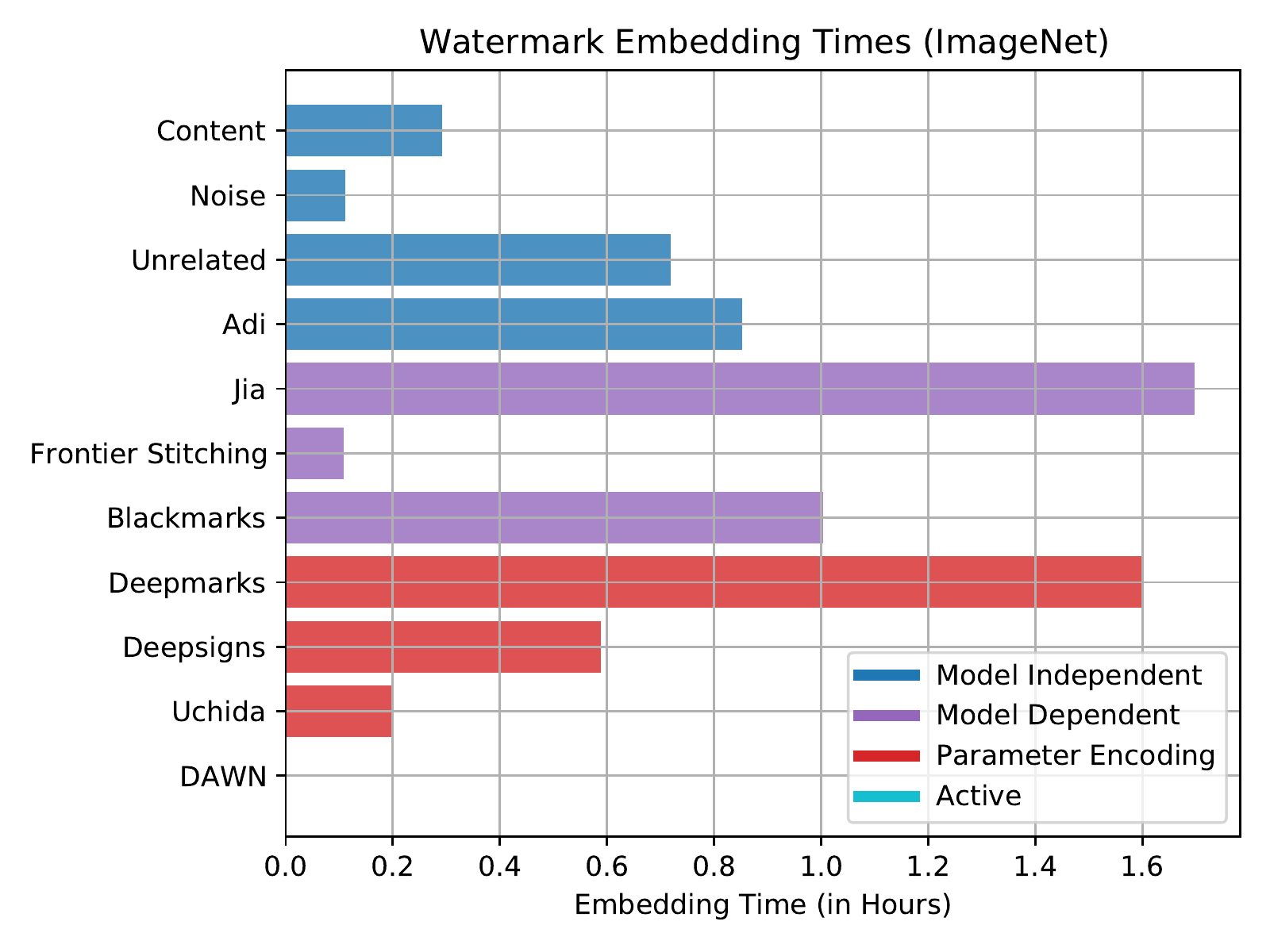}}\label{fig:imagenet_embedding_time}}
	\subfloat[]{\raisebox{-0.5\height}{\includegraphics[width=.33\linewidth]{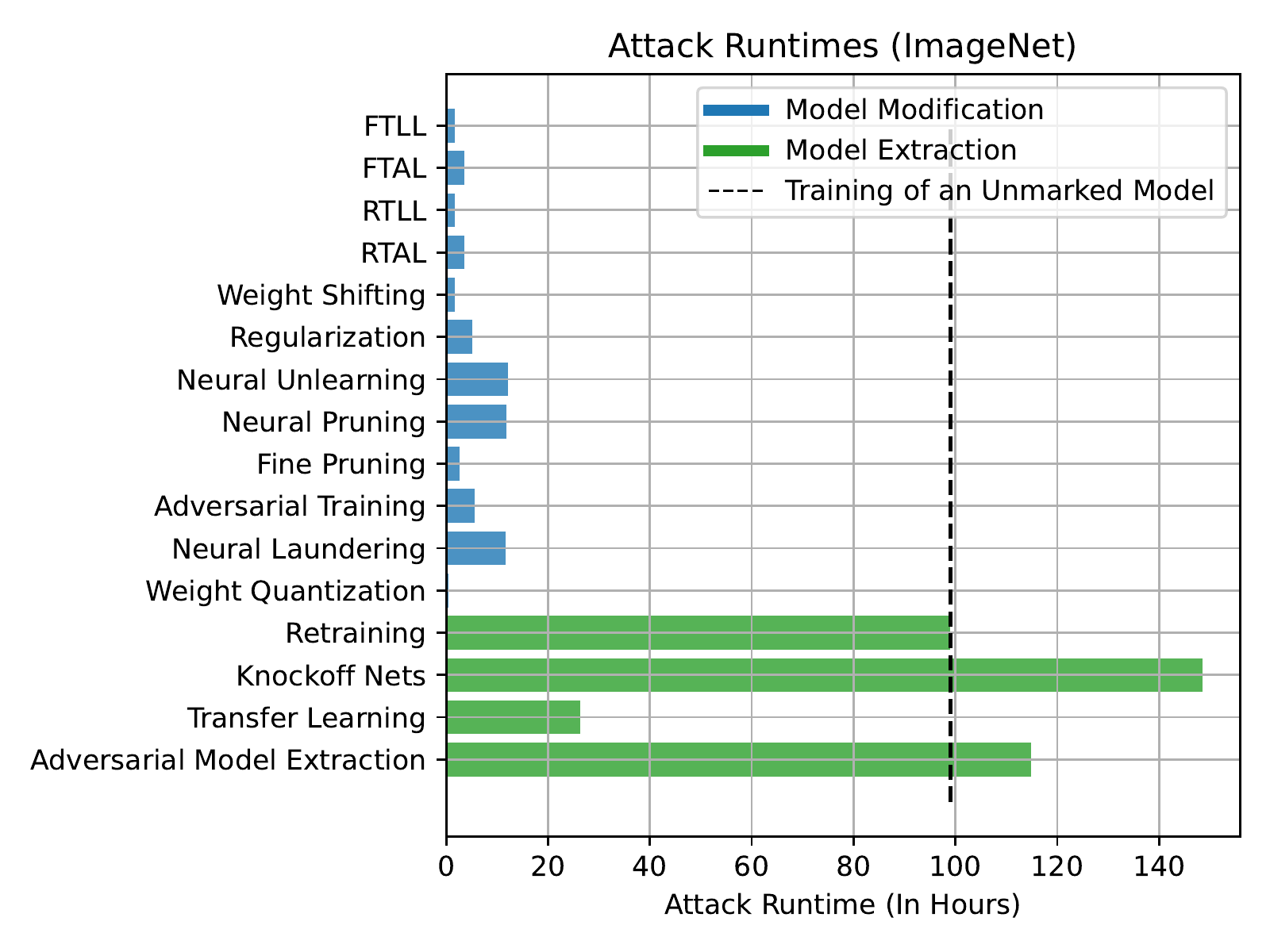}}\label{fig:imagenet_attack_time}}
	\subfloat[]{\raisebox{-0.5\height}{\includegraphics[width=.33\linewidth]{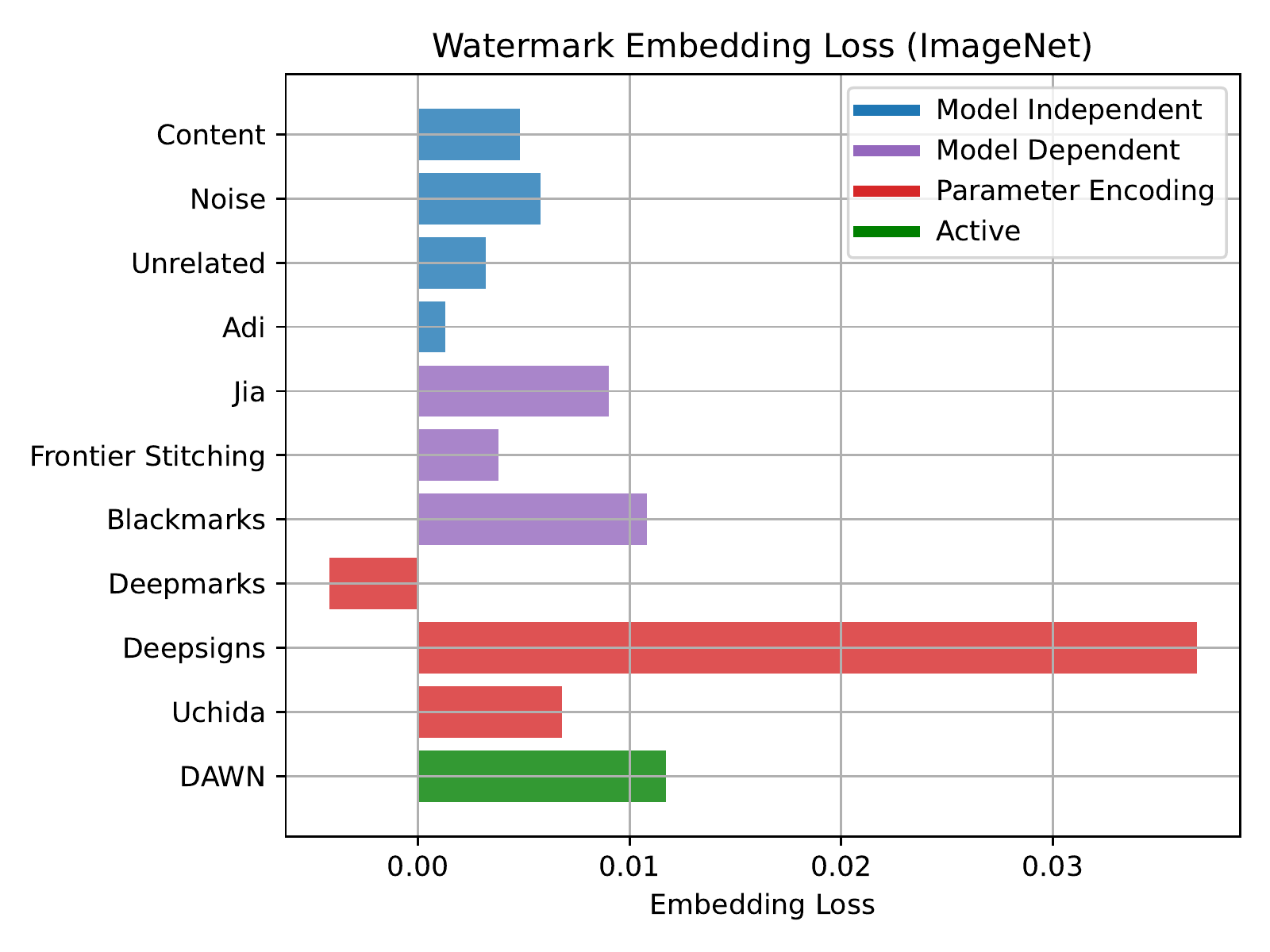}}\label{fig:imagenet_embedding_loss}}
    \caption{The measured runtimes for embedding and attacking a watermark and the embedding losses for each watermark on CIFAR-10 (top) and ImageNet (bottom).
    Figures~\ref{fig:cifar_embedding_time}, \ref{fig:imagenet_embedding_time} show the embedding times and Figures~\ref{fig:cifar_attack_time}, \ref{fig:imagenet_attack_time} show the removal attack runtime.
    Figures~\ref{fig:cifar_embedding_loss}, \ref{fig:imagenet_embedding_loss} show the embedding loss of each watermarking scheme, which is the difference in test accuracy between an unmarked model and the (marked) source model.\label{fig:defense_statistics} }
\end{figure*}

\subsection{Setup}
We implement all watermark schemes and removal attacks in our novel Watermark-Robustness-Toolbox (WRT) with PyTorch~\cite{NEURIPS2019_9015} running as its backend.
WRT will be made available as open-source code, which allows independently verifying our empirical results.
All reported runtimes in this paper were obtained using (single) Tesla P100 GPUs. 

\subsection{Datasets}
We embed watermarks into source models trained on the image classification datasets CIFAR-10~\cite{cifar10} and ImageNet~\cite{imagenet}. 
The Open Images~\cite{openimages} dataset is used as a transfer dataset (see Section~\ref{sec:attacker_capabilities}). 
Our method of splitting the dataset between the attacker and defender differs depending on the attack's category. 
For model modification attacks, the attacker has access to a third of the dataset and the defender can access the remaining two thirds.
Model extraction attacks require more data to achieve a high test accuracy, hence the attacker and defender have access to the entire training dataset. 
We refer to Appendix~\ref{sec:datasets} for a description of the datasets and details on our method of splitting the training dataset.

\subsection{Model Architectures}
All of our experiments assume that the attacker knows the source model's architecture. 
For CIFAR-10, we use the wide ResNet 28x10~\cite{zagoruyko2016wide} and for ImageNet the ResNet-50~\cite{he2016deep} architectures. 
We also perform cross-architecture retraining using a DenseNet-121~\cite{huang2017densely} for CIFAR-10 and ImageNet.

\subsection{Runtimes and Embedding Losses}
\label{sec:experiments:embedding_loss}
We report the runtimes for the removal attacks and watermark embeddings. 
Since the runtimes are influenced by the choice of parameters, the results can only show general trends. 
We ensured choosing parameters and training configurations that an attacker or defender would also likely choose in practice, such as using early stopping for the embedding. 
For a detailed description of the chosen parameters and implementation details we refer to Appendixes~\ref{sec:schemes} and \ref{sec:attacks}.
\Cref{fig:cifar_embedding_time,fig:cifar_attack_time,fig:cifar_embedding_loss} show results for CIFAR-10 and \Cref{fig:imagenet_embedding_time,fig:imagenet_attack_time,fig:imagenet_embedding_loss} for ImageNet. 
All graphs are shown as horizontal bar charts with the watermarking scheme or removal attack on the y-axis and the runtime or the embedding loss on the x-axis.
The coloring indicates the category of a scheme or removal attack. 

\textbf{Embedding Runtimes.} 
\Cref{fig:cifar_embedding_time,fig:imagenet_embedding_time} show the embedding runtimes for CIFAR-10 and ImageNet. 
We refer to the \emph{training time} as the time it takes to train an unmarked model from scratch.
This training time serves as a point of reference to assess the practicality of removal attacks and watermarking schemes. 
For CIFAR-10 and ImageNet we observe a training time of 1h and 100h, respectively. 

On CIFAR-10, model independent schemes have the highest embedding time of about 20\% of the training time, whereas 
parameter encoding schemes have the lowest embedding times and require only about 9\% of the training time. 
We do not consider the runtime for the active scheme DAWN but point out that deploying DAWN incurs computational costs for each inference. 
On ImageNet, we observe that schemes such as Jia and Deepmarks require considerably more time than on CIFAR-10, whereas model independent schemes are relatively fast to embed. 
The longest embedding time has Jia with more than 1.6\% of the training time. 
These embedding times are low compared to the training times for both datasets, and we conclude that all surveyed schemes are efficient. 

\textbf{Attack Runtimes.} 
\Cref{fig:cifar_attack_time,fig:imagenet_attack_time} show the attack runtimes for CIFAR-10 and ImageNet. 
Input Preprocessing attacks are not shown, because they run only during inference.
We observe that the runtimes of all attacks are proportionally similar on CIFAR-10 and ImageNet. 
On both datasets, model extraction attacks require significantly longer than model modification attacks. 
Transfer learning is an exception for a model extraction attack that is relatively fast as it requires about $40\%$ of the training time on CIFAR-10 and roughly $25\%$ of the training time on ImageNet. 
Knockoff is the slowest attack which takes considerably longer than retraining due to the larger size of the training dataset. 

\textbf{Embedding Losses.} \Cref{fig:cifar_embedding_loss,fig:imagenet_embedding_loss} show the embedding loss for each scheme, which is the drop in test accuracy due to embedding the watermark into the source model (see Section~\ref{sec:measurements}).
Embedding losses for CIFAR-10 and ImageNet are about one percentage point, with the exception of Deepsigns on ImageNet, which has an embedding loss of more than three percentage points. 
The parameter encoding scheme Deepmarks incurs the lowest embedding loss on both datasets. 
\begin{figure*}
	\centering
	\subfloat[]{\raisebox{-0.5\height}{\includegraphics[trim=0 0 0 0.5cm,width=.33\linewidth]{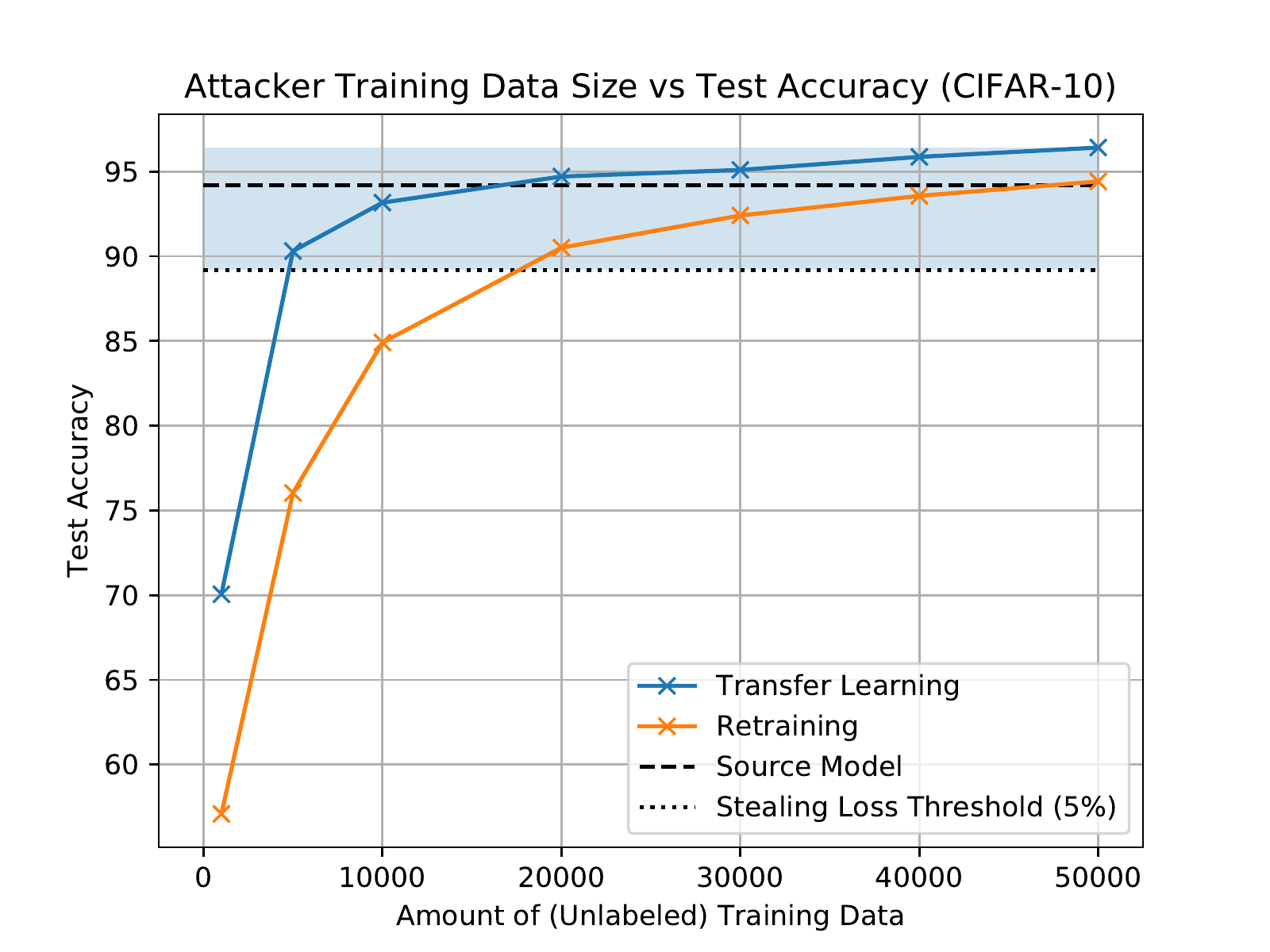}}\label{fig:cifar_lineplot}}
	\subfloat[]{\raisebox{-0.5\height}{\includegraphics[width=.32\linewidth]{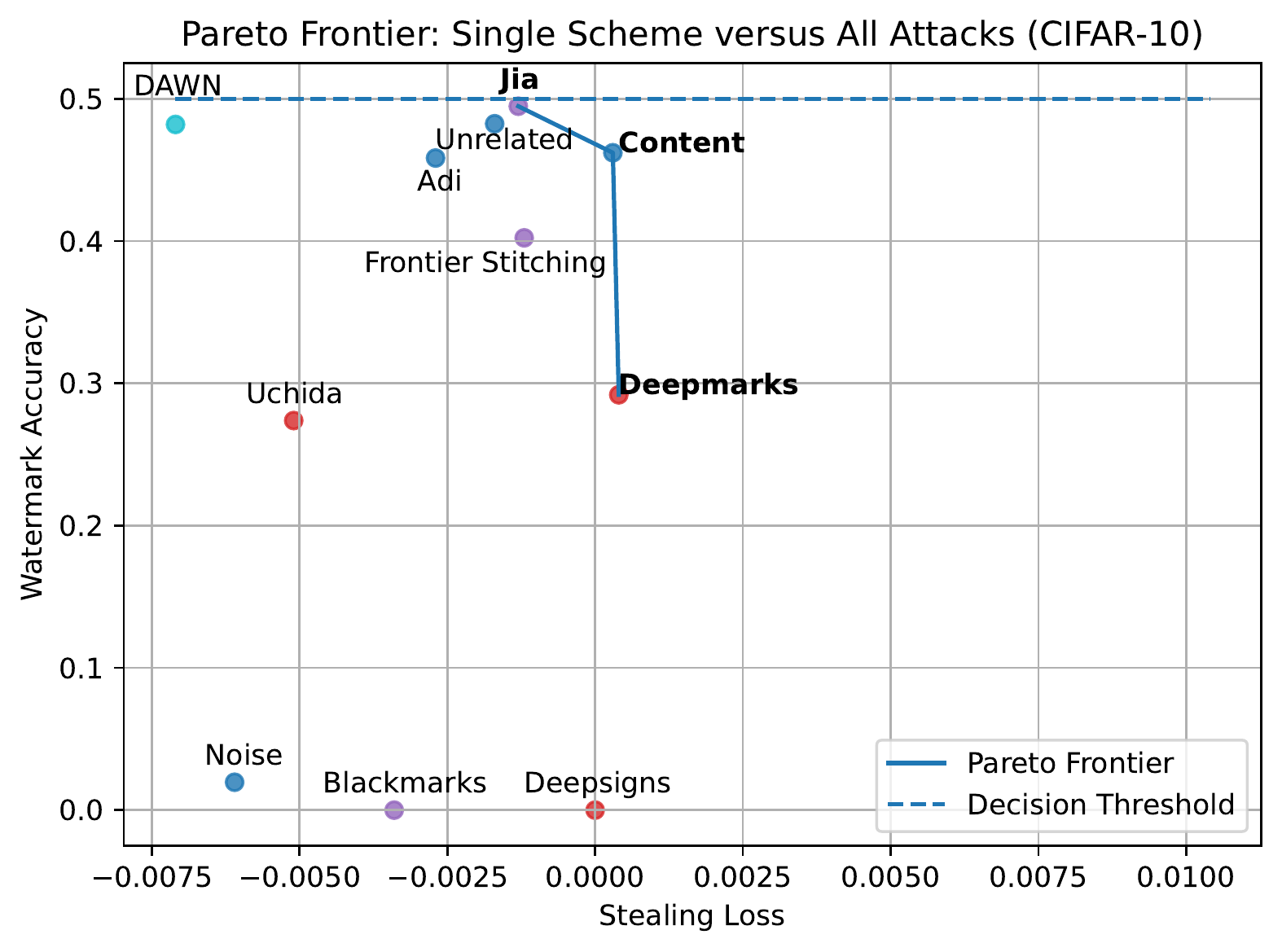}}\label{fig:cifar_paretodefense}}
	\subfloat[]{\raisebox{-0.5\height}{\includegraphics[width=.33\linewidth]{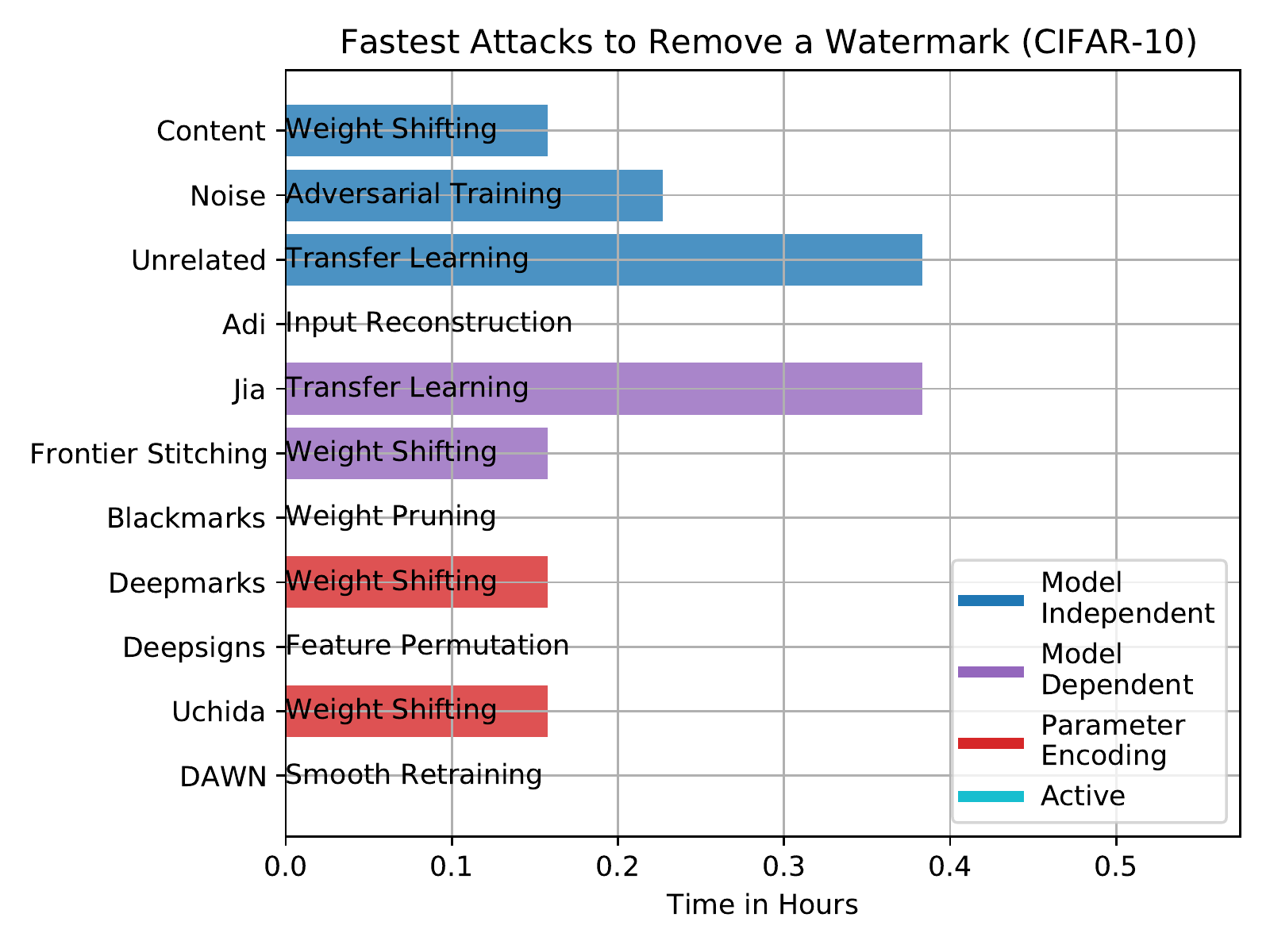}}\label{fig:cifar_fastestattack}} \\
	
	\subfloat[]{\raisebox{-0.5\height}{\includegraphics[trim=0 0 0 0.5cm, width=.33\linewidth]{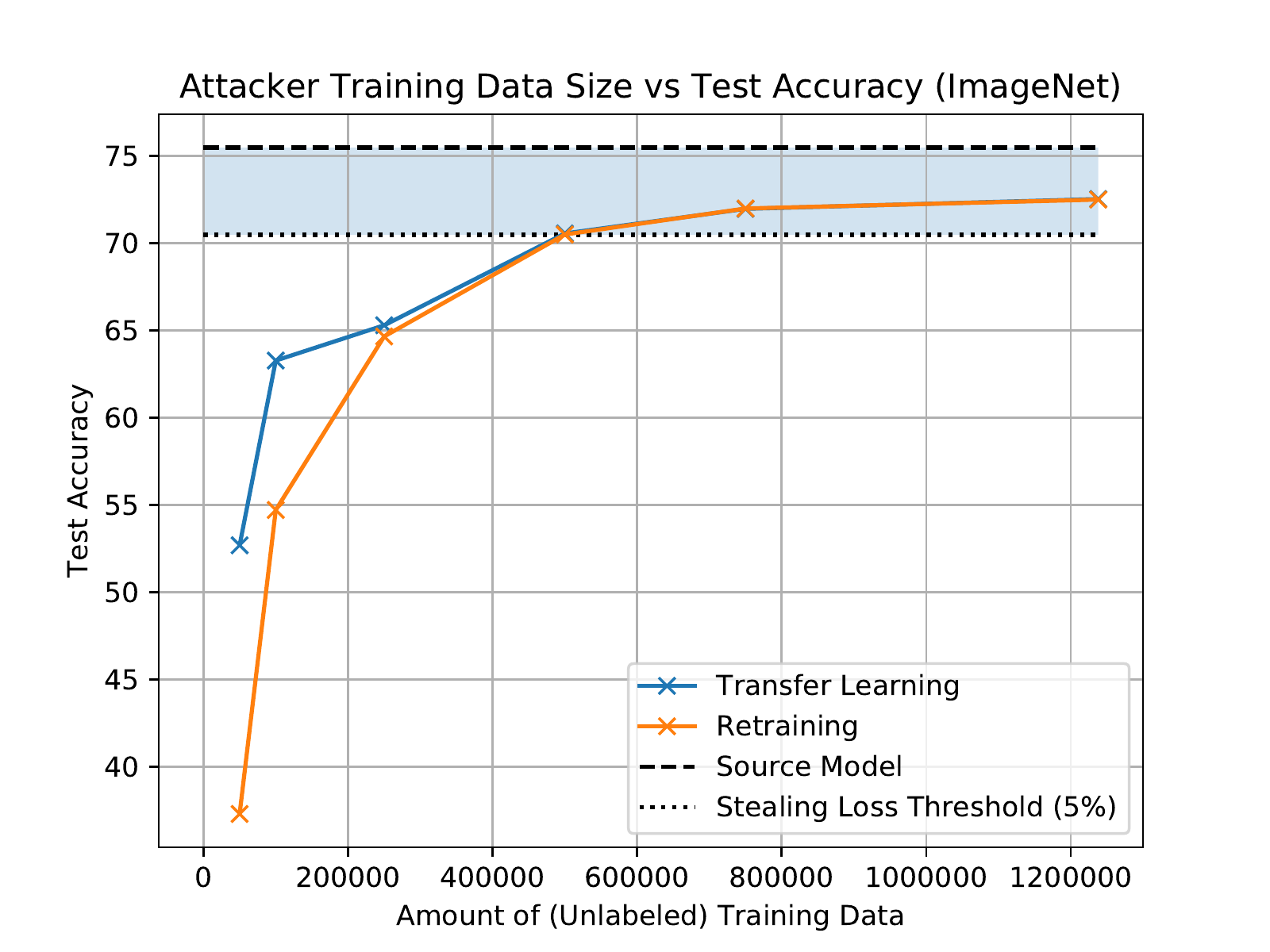}}\label{fig:imagenet_lineplot}}
	\subfloat[]{\raisebox{-0.5\height}{\includegraphics[width=.32\linewidth]{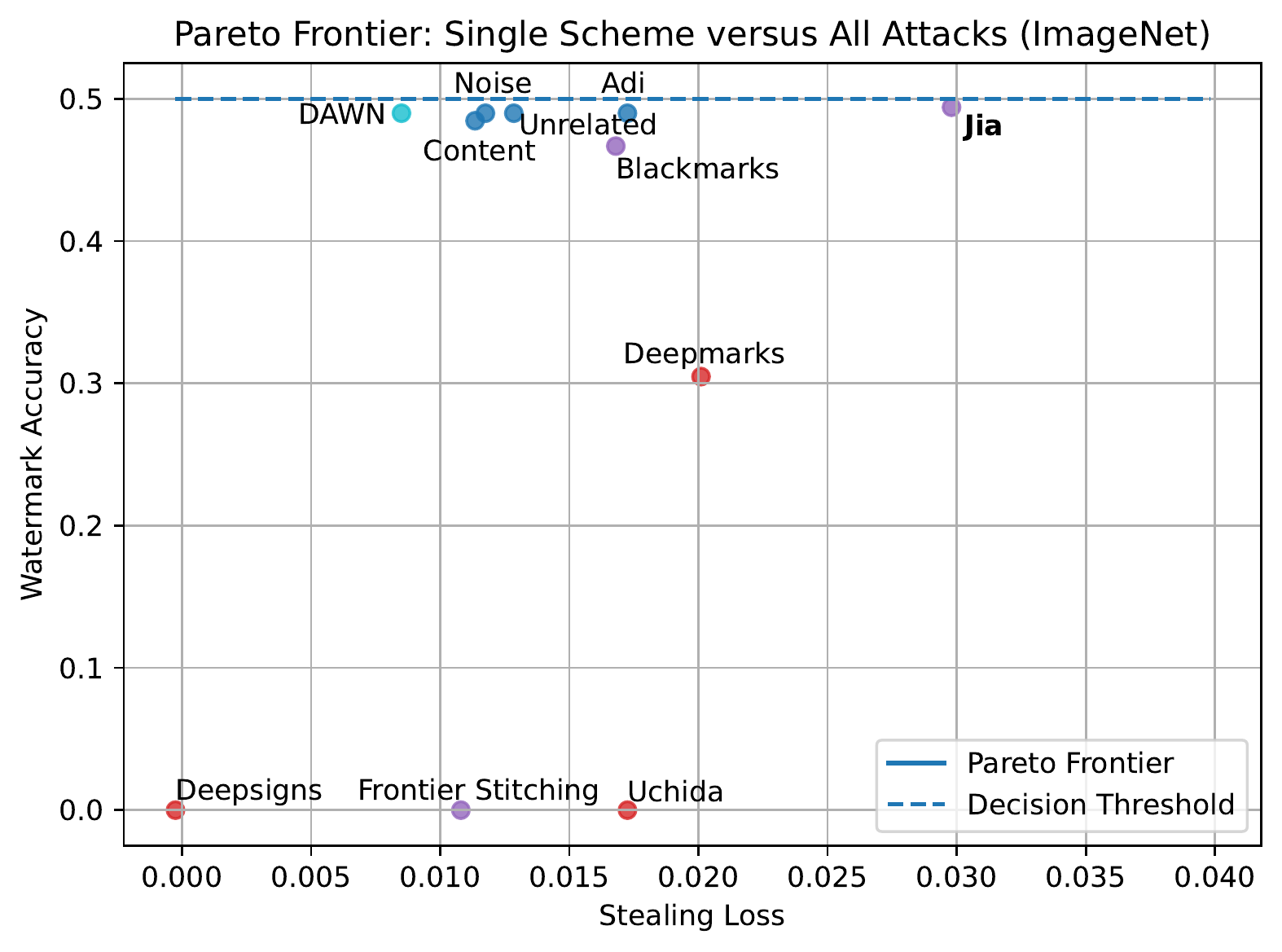}}\label{fig:imagenet_paretodefense}}
	\subfloat[]{\raisebox{-0.5\height}{\includegraphics[width=.33\linewidth]{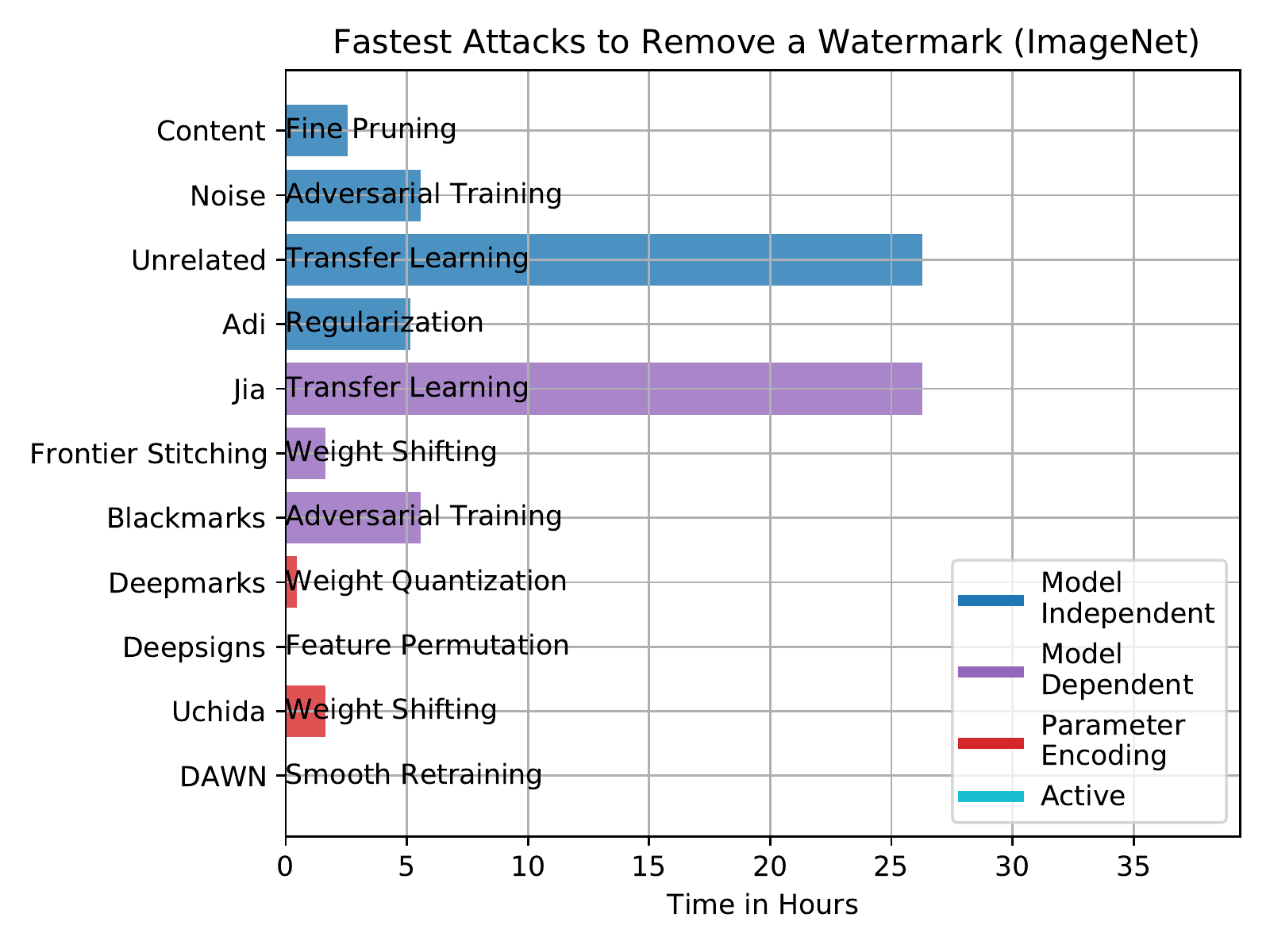}}\label{fig:imagenet_fastestattack}}
	\caption{
	Figures (a,d) compare the amount of training data required for the transfer learning and retraining attacks to achieve a given test accuracy. 
	Figures (b,e) show the Pareto frontier for all watermarking schemes with respect to the stealing loss~(defined in Section~\ref{sec:measurements}) and watermark accuracy of the best attack. 
	A watermark accuracy lower than $\theta'=0.5$ means that the watermark is not robust. 
	Figures (c, f) show the fastest attack that removes each watermark. 
	For DAWN, the attacker has to obtain white-box access by extracting the source model before using other attacks. For a fair comparison with other schemes, we do not consider this extraction runtime.  \label{fig:single_results} }
\end{figure*}
\subsection{Robustness of Watermarking Schemes}
\label{sec:exp_robustness}
In this section, we analyze the robustness of each watermarking scheme against all attacks. 
This means that the defender can choose from a set of parameters for a single watermarking scheme, whereas an attacker can choose from all parameters for all removal attacks. 
The goal of this analysis is to evaluate whether any watermarking scheme can be considered robust against an adaptive adversary.
We assume that the attacker knows the watermarking scheme chosen by the defender but not its parameters. 

\textbf{Robustness.} The results are illustrated in \Cref{fig:cifar_paretodefense,fig:imagenet_paretodefense} in the form of a scatter plot.  
The x-axis shows the stealing loss, which is the drop in test accuracy in the surrogate model compared to the source model, and the y-axis shows the rescaled watermark accuracy (see Section~\ref{sec:measurements}).
A watermark accuracy lower than $\theta'=0.5$ means that the watermark has been removed. 
We highlight $\theta'$ by a dashed line in the graph.
We draw the \emph{Pareto frontier}, which is the set of watermarking schemes with a watermark accuracy or stealing loss so that no other watermarking scheme improves upon both metrics. 
Jia, Content, and Deepmarks are members of the Pareto frontier for CIFAR-10 and only Jia for ImageNet. 

We observe that none of the watermarking schemes is robust. 
For CIFAR-10, the marked source models can be stolen with a stealing loss of less than one percentage point, i.e., without a considerable loss of utility. 
For ImageNet, we observe that removal attacks incur a higher stealing loss overall.
Jia has the highest stealing loss of three percentage points, whereas the remaining watermarking schemes have a stealing loss of at most two percentage points. 
We designed a set of adaptive attacks against a subset of watermarking schemes and feature their results separately as following.  
We refer to Appendix~\ref{sec:attacks} for a detailed description of all attacks. 
\begin{itemize}
    \item \textbf{Smooth Retraining}: The smooth retraining attack is adapted to the active watermarking scheme DAWN. 
    The idea is to query DAWN multiple times with the same image, using a different affine transformation (e.g., cropping, horizontal flipping) for each query. 
    The label for each image is the mean over all received labels for each image. 
    Smooth retraining is the only attack that removes DAWN on CIFAR-10. 
    \item \textbf{Feature Permutation}: Hidden layer neurons are permutation invariant, meaning that we can apply a random permutation on the features without losing any utility of the model. 
    We observe that Deepsigns is the only scheme that is not robust against feature permutation attacks.  
    \item \textbf{Weight Shifting}: Weight shifting perturbs the filter weights of each convolutional layer by the negative mean over all its filters, adds a small amount of noise, and fine-tunes the model. 
    We observe that weight shifting is the only model modification attack that removes Uchida on CIFAR-10 and ImageNet. 
\end{itemize}

\textbf{Fastest Attacks.} \Cref{fig:cifar_fastestattack,fig:imagenet_fastestattack} show the fastest attacks that successfully remove a watermark. 
On CIFAR-10, we observe that some schemes such as Deepsigns, Blackmarks, and Adi can be removed with a negligible runtime, whereas Jia and Unrelated require the highest runtime. 
On ImageNet, we observe that the removal of the watermarks from Unrelated and Jia requires the highest runtime, whereas parameter encoding schemes can be removed in the shortest amount of time. 
For both datasets, we observe that the fastest attacks depend on the watermarking scheme, i.e., there is no single fastest attack or attack category against all watermarking schemes. 

\textbf{Dataset Availability.} We stated that the dataset available to a model extraction attack is larger than for model modification attacks. 
We ablate over the amount of data available to the attacker to achieve a given test accuracy. 
This is relevant to discuss the practicality of model extraction attacks because the attacker wants to minimize both (i) the training time and (ii) the amount of data required to perform an attack.

\Cref{fig:cifar_lineplot,fig:imagenet_lineplot} show the amount of unlabeled data in relation to the surrogate model's test accuracy for CIFAR-10 and ImageNet. 
The attacker trains their surrogate model on data labeled by source models with a test accuracy of 94.20\% on CIFAR-10 and 75.48\% on ImageNet. 
On CIFAR-10, we observe that transfer learning achieves a significantly higher test accuracy than retraining from scratch using the same amount of data. 
Retraining requires at least about 20k samples to perform a successful attack, whereas transfer learning needs only about 5k samples. 
On ImageNet, the difference between retraining and transfer learning goes to zero when more than 250k samples are available to the attacker. 
Performing a successful removal attack requires at least 500k samples. 
While transfer learning still requires the same amount of data as retraining from scratch, we point out that transfer learning requires significantly less computation time. 

\begin{figure*}
    \captionsetup[subfigure]{labelformat=empty}
	\centering
	\subfloat[]{\raisebox{-0.5\height}{\includegraphics[width=0.80\linewidth]{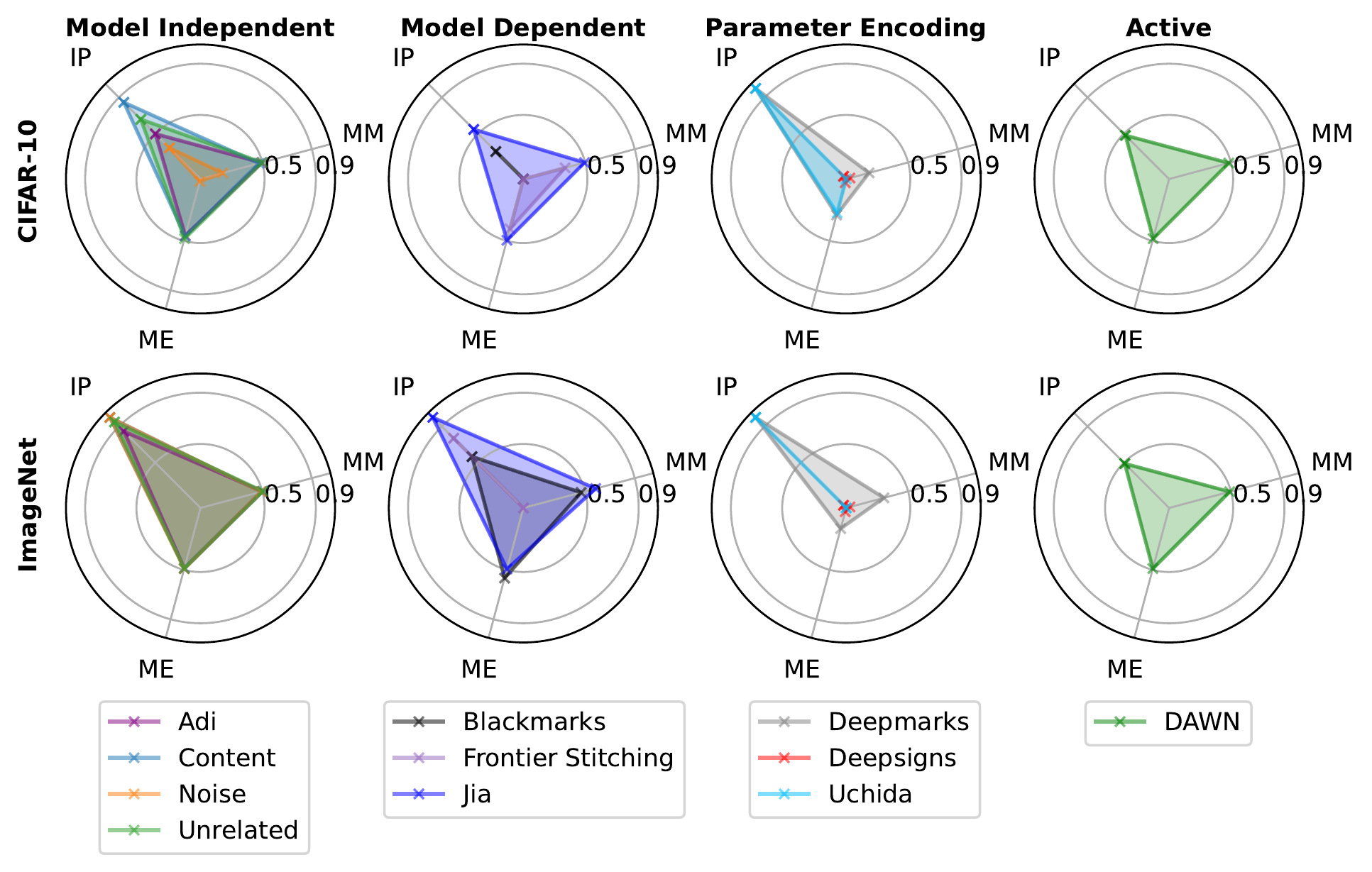}}\label{fig:radarplot}}
	\hspace{5ex}
	\caption{
	This figure illustrates the robustness of each surveyed watermarking scheme against categories of attacks for CIFAR-10 (top) and ImageNet (bottom). 
	The axes show the (scaled) watermark accuracy of a scheme against the best attack from each category. 
	A watermark is robust against a category if the watermark accuracy is at least $\theta'=0.5$. 
	The scheme and attack parameters are chosen using the Nash Equilibrium, and we ignore attacks when their stealing loss exceeds five percentage points. 
	The attack categories are Input Preprocessing (IP), Model Modification (MM), and Model Extraction (ME). }
	\label{fig:radarplots}
\end{figure*}

\subsection{Robustness against Attack Categories}

In the previous section, we showed that none of the watermarking schemes is robust against all attacks. 
We further analyze the robustness of each watermarking scheme against categories of removal attacks. 
The defender can choose from the set of parameters for each watermarking scheme, and the attacker can choose from the set of parameters for attacks of only one category. 
This analysis provides insights into the vulnerability of watermarking schemes to certain attack categories.
We refer the reader to Table~\ref{tab:removal_attacks} for a list of all attacks and their categories. 

Figure~\ref{fig:radarplots} shows a radar plot of our result for CIFAR-10 and ImageNet. 
The radar plot axis shows the watermark accuracy of each scheme against the best, successful attack from each attack category. 
A larger covered area of the watermarking scheme in the plot illustrates higher robustness to multiple attack categories.
A scheme is robust against the attack category if the watermark accuracy is at least $\theta'=0.5$ (see Section~\ref{sec:measurements}).
We analyze the results for each category. 

\textbf{Input Preprocessing.}
We observe that input preprocessing attacks often do not remove a watermark on either CIFAR-10 or ImageNet, but these attacks often impact the watermark accuracy. 
Input smoothing and input reconstruction are effective against Adi and Noise on CIFAR-10, but not on ImageNet. 
We always apply feature permutation because it does not impact the model's utility and requires negligible computational costs. 
For this reason, Deepsigns, which is vulnerable to feature permutation, is removed by input preprocessing attacks for both CIFAR-10 and ImageNet. 
Similarly, DAWN is not robust because it requires extracting a surrogate model prior to running an input preprocessing or model modification attack.
We extract a surrogate model for DAWN using smooth retraining, which already removes the watermark.

\textbf{Model Modification.}
Model modification attacks are successful at removing all watermarks for CIFAR-10 and ImageNet, except for Jia on ImageNet. 
Many surveyed watermarking schemes are vulnerable against multiple model modification attacks, whereas other schemes such as Uchida are only vulnerable to our adaptive weight shifting attack. 
Similar to input preprocessing attacks, we observe that model modification attacks that do not remove the watermark can still significantly lower the watermark accuracy.

\textbf{Model Extraction.}
We observe that almost none of the schemes is robust to model extraction attacks on CIFAR-10 and ImageNet. 
The most effective attack is transfer learning for both CIFAR-10 and ImageNet because it requires a fraction of the training time for an unmarked model, and it removes almost all of the surveyed watermarks. 
Notable exceptions are Noise and Blackmarks, which are robust against transfer learning on ImageNet, but Noise is not robust against retraining on ImageNet and Blackmarks is not robust against adversarial training. 
Retraining, distillation, and adversarial training from scratch yield similar results as transfer learning, but they require (i) at least as much data and (ii) have a significantly longer runtime.
Therefore we do not evaluate distillation and adversarial model extraction on ImageNet if a model is already vulnerable to transfer learning or retraining.

In summary, we conclude that model extraction attacks are the most effective removal attacks against a majority of watermarks. 
Jia and Blackmarks are robust against retraining, but Jia is not robust against transfer learning, and Blackmarks is not robust against adversarial training. 
Even when a scheme is robust to retraining with the same architecture, the attacker can obtain a well-trained surrogate model by switching to a different architecture. 
We believe that transfer learning is more effective at removing some watermarks because the model re-uses low-level features learned from another task. Hence, watermarks encoded into low-level features are less likely to be robust against transfer learning. 
None of the parameter encoding schemes is robust to transfer learning, also because extraction of such a watermark is not defined for a different model architecture. 
For example, Uchida defines a secret watermarking key that expects a layer's weights to be in the same shape as the source model's layer used for the embedding.
Input preprocessing attacks are often non-successful at removing a watermark, but they can reduce the watermark accuracy. 
Model modification attacks, especially our novel adaptive attacks, are successful in removing the watermark of a subset of watermarking schemes and require (i) significantly fewer data and (ii) computational resources than model extraction attacks. 

\begin{table*}[]
    \centering
    \begin{tabular}{|l|ccccccccccc|}
\hline
\diagbox{\textbf{Attack}}{\textbf{Watermark}}
&  \makecell{\textbf{Content}\\\cite{zhang2018protecting}} & \makecell{\textbf{Noise}\\\cite{zhang2018protecting}} & \makecell{\textbf{Unrelated}\\\cite{zhang2018protecting}} & \makecell{\textbf{Adi}\\\cite{adi2018turning}} & \makecell{\textbf{Jia}\\\cite{jia2020entangled}} & \makecell{\textbf{FS}\\\cite{le2020adversarial}} & \makecell{\textbf{BM}\\\cite{chen2019blackmarks}} & \makecell{\textbf{Deepmarks}\\\cite{chen2018deepmarks}} & \makecell{\textbf{Deepsigns}\\\cite{rouhani2018deepsigns}} & \makecell{\textbf{Uchida}\\\cite{uchida2017embedding}} & \makecell{\textbf{DAWN}\\\cite{szyller2019dawn}} \\ \hline

 \multicolumn{1}{|c|}{\small \textsc{Input Preprocessing}} &&&&&&&&&&& \\
 
 \multicolumn{1}{|c|}{\makecell{Input  Smoothing~\cite{xu2017feature} \\ (Gaussian Kernel)}}  &        
 \checkmark/\checkmark &  % Content
 \checkmark/\checkmark &  % Noise
 \checkmark/\checkmark &  % Unrelated
 \checkmark/\checkmark &  % Adi
 \checkmark/\checkmark &  % Jia
 \xmark/\checkmark &  % FS
 \checkmark/\checkmark &  % Blackmarks
 \checkmark/\checkmark &  % Deepmarks
 \checkmark/\checkmark &  % Deepsigns
 \checkmark/\checkmark &  % Uchida
 \checkmark/\checkmark    % DAWN
 \\ \hline
 
\multicolumn{1}{|c|}{\small \textsc{Model Modification}} &&&&&&&&&&& \\
 
\multicolumn{1}{|c|}{Regularization~\cite{shafieinejad2019robustness}}  &        
 \xmark/\checkmark &  % Content
 \checkmark/\checkmark &  % Noise
 \checkmark/\checkmark &  % Unrelated
 \xmark/\xmark &  % Adi
 \xmark/\checkmark &  % Jia
 \xmark/\checkmark &  % FS
 \xmark/\checkmark &  % Blackmarks
 \checkmark/\checkmark &  % Deepmarks
 \checkmark/\checkmark &  % Deepsigns
 \checkmark/\checkmark &  % Uchida
 \checkmark/\xmark    % DAWN
 \\ 

 \multicolumn{1}{|c|}{\makecell{Neural Cleanse~\cite{wang2019neural} \\ (Unlearning)}}  &        
 \checkmark/\checkmark &  % Content
 \checkmark/\checkmark &  % Noise
 \checkmark/\checkmark &  % Unrelated
 \checkmark/\checkmark &  % Adi
 \checkmark/\checkmark &  % Jia
 \checkmark/\checkmark &  % FS
 \checkmark/\checkmark &  % Blackmarks
 \checkmark/\checkmark &  % Deepmarks
 \checkmark/\checkmark &  % Deepsigns
 \checkmark/\checkmark &  % Uchida
 \checkmark/\checkmark    % DAWN
 \\ 
 
 \multicolumn{1}{|c|}{Feature Permutation~(Ours)}  &        
 \checkmark/\checkmark &  % Content
 \checkmark/\checkmark &  % Noise
 \checkmark/\checkmark &  % Unrelated
 \checkmark/\checkmark &  % Adi
 \checkmark/\checkmark &  % Jia
 \checkmark/\checkmark &  % FS
 \checkmark/\checkmark &  % Blackmarks
 \checkmark/\checkmark &  % Deepmarks
 \xmark/\xmark &  % Deepsigns
 \checkmark/\checkmark &  % Uchida
 \checkmark/\checkmark    % DAWN
 \\
 
   \multicolumn{1}{|c|}{Weight Shifting~(Ours)}  &        
 \xmark/\checkmark &  % Content
 \checkmark/\checkmark &  % Noise
 \xmark/\checkmark &  % Unrelated
 \xmark/\checkmark &  % Adi
 \checkmark/\checkmark &  % Jia
 \checkmark/\xmark &  % FS
 \xmark/\checkmark &  % Blackmarks
 \xmark/\xmark &  % Deepmarks
 \xmark/\xmark &  % Deepsigns
 \xmark/\xmark &  % Uchida
 \checkmark/\checkmark    % DAWN
 \\  \hline

 \multicolumn{1}{|c|}{\small \textsc{Model Extraction}} &&&&&&&&&&& \\

  \multicolumn{1}{|c|}{Knockoff Nets~\cite{orekondy2019knockoff}}  &        
 \checkmark/\checkmark &  % Content
 \checkmark/\checkmark &  % Noise
 \checkmark/\checkmark &  % Unrelated
 \checkmark/\checkmark &  % Adi
 \checkmark/\checkmark &  % Jia
 \checkmark/\checkmark &  % FS
 \checkmark/\checkmark&  % Blackmarks
 \xmark/\checkmark &  % Deepmarks
 \xmark/\checkmark &  % Deepsigns
 \xmark/\checkmark &  % Uchida
 -    % DAWN
 \\ 
  \multicolumn{1}{|c|}{Retraining~\cite{tramer2016stealing}}  &        
 \xmark/\xmark &  % Content
 \xmark/\xmark &  % Noise
 \xmark/\xmark &  % Unrelated
 \xmark/\xmark &  % Adi
 \checkmark/\checkmark &  % Jia
 \xmark/\checkmark &  % FS
 \xmark/\xmark &  % Blackmarks
 \xmark/\xmark &  % Deepmarks
 \xmark/\xmark &  % Deepsigns
 \xmark/\xmark &  % Uchida
 \checkmark/\xmark    % DAWN
 \\ 
 
  \multicolumn{1}{|c|}{Smooth Retraining~(Ours)}  &        
 - &  % Content
 - &  % Noise
 - &  % Unrelated
 - &  % Adi
 - &  % Jia
 - &  % FS
 - &  % Blackmarks
 - &  % Deepmarks
 - &  % Deepsigns
 - &  % Uchida
 \xmark/\xmark    % DAWN
 \\ 
 
 \multicolumn{1}{|c|}{\makecell{Cross-Architecture \\Retraining}}  &        
 \xmark/\xmark &  % Content
 \xmark/\xmark &  % Noise
 \xmark/\xmark &  % Unrelated
 \xmark/\xmark &  % Adi
 \xmark/\checkmark &  % Jia
 \xmark/\checkmark &  % FS
 \xmark/\checkmark&  % Blackmarks
 \xmark/\xmark &  % Deepmarks
 \xmark/\xmark &  % Deepsigns
 \xmark/\xmark &  % Uchida
 \checkmark/\xmark    % DAWN
 \\
 
 \multicolumn{1}{|c|}{Transfer Learning~\cite{torrey2010transfer}}  &        
 \xmark/\xmark &  % Content
 \xmark/\checkmark &  % Noise
 \xmark/\xmark &  % Unrelated
 \xmark/\xmark &  % Adi
 \xmark/\xmark &  % Jia
 \xmark/\xmark &  % FS
 \xmark/\checkmark &  % Blackmarks
 \xmark/\xmark &  % Deepmarks
 \xmark/\xmark &  % Deepsigns
 \xmark/\xmark &  % Uchida
 \checkmark/\xmark    % DAWN
 \\ \hline
 
\end{tabular}
    \caption{A summary of the robustness for each watermarking scheme against selected attacks. A checkmark ('\checkmark') indicates that the scheme is robust, whereas a cross ('\xmark') indicates that the scheme is \emph{not} robust to this attack. A dash indicates that the attack has not performed against the watermarking scheme (e.g., because it is an adaptive attack designed against a subset of schemes). By two consecutive marks, we indicate the robustness on CIFAR-10 and ImageNet.
    \label{tab:single_attack_vs_defense}}
\end{table*}

\subsection{Attack's Effectiveness.}
\Cref{tab:single_attack_vs_defense} shows whether a scheme is robust against an attack on CIFAR-10 and ImageNet for a subset of attacks. 
We make the observations that (i) attacks designed against one category of watermarks are not necessarily effective against watermarks from this category, and (ii) no scheme is robust against all model extraction attacks. 
Neural Cleanse~\cite{aiken2020neural} and Regularization~\cite{shafieinejad2019robustness} were designed against model independent watermarks, but they often only decrease the watermark accuracy instead of removing the watermark. 
Jia is robust against retraining, but not against transfer learning suggesting that it is encoded into the low-level features of the source model. 
Transfer learning does not re-learn these low-level features from scratch, which could explain why transfer learning is more effective than retraining at removing the Jia watermark. 

\subsection{Dominant Attacks}
This section analyzes whether a \emph{dominant} attack exists that removes all watermarks.
The existence of a dominant attack would mean that an attacker does not require knowledge about the scheme used by the defender to remove their watermark. 
The attacker can choose from the set of parameters for a single attack, whereas the defender can choose from the set of parameters for all watermarking schemes.
We observe that transfer learning is dominant for CIFAR-10, but there exists no dominant attack for ImageNet.

\textbf{Creating Dominant Attacks.}
We now evaluate whether it is possible to find \emph{combined} attacks that are dominant for source models trained on ImageNet. 
A combined attack performs many attacks in sequence. 
Our empirical results show that transfer learning combined with label smoothing is a dominant attack that removes all eleven watermarks on CIFAR-10 and ImageNet. 
The threat of combined attacks to the robustness of watermarking schemes has not yet been explored, and we show that combined attacks can pose a significant threat.  

\section{Discussion}
In this section, we discuss the practicality of the evaluated removal attacks and argue that they are real-world threats to DNN watermarking. 
We identify three requirements for the attacker: (1) computational resources, (2) dataset availability, and (3) pre-trained models for transfer learning. 
Then we present guidelines for designing future watermarking schemes and discuss the implications of our work for future research. 

\textbf{Computational Resources.}
Related work often restricts the availability of computational resources to the attacker in their threat model~\cite{uchida2017embedding, adi2018turning, chen2018deepmarks} and claims robustness against attackers with limited computational resources. 
We believe that this assumption is not realistic and that a motivated attacker is not limited by computational resources. 
While it may be the adversary's objective to minimize computational resources, there is no theoretical guarantee that the adversary's learning problem will be a hard instance and require infeasible resources in some security parameters.
%We believe that this assumption is not realistic and that a motivated attacker is not limited by computational resources. 
Quite to the contrary, for the classification problems considered in this paper, the adversary's costs are very feasible.
Using shared GPUs in the cloud, the monetary costs are proportional to the attack's runtime.
All runtimes in our paper were obtained on (single) Tesla P100 GPUs, which incur a cost of $0.43\$$ per on-demand hour of GPU-time\footnote{\url{https://cloud.google.com/compute/gpus-pricing}}. 
Training a ResNet-50 model from scratch on ImageNet, consisting of 1.28 million images, takes about 100 hours and costs 43\$.
Transfer learning a model takes only 23 hours and brings down the costs to about 10\$. 
There are even more optimized implementations~\cite{coleman2017dawnbench} than ours, which achieve lower costs through various optimizations, e.g., by training on multiple GPUs, utilizing TPUs, or choosing more efficient model architectures.
We conclude that in absolute terms, the price for computational resources is almost insignificant and is likely not a deterrent for the attacker.
%We conclude that in absolute terms, the price for computational resources is almost insignificant and does not deter a motivated attacker even for large ImageNet classification models.

\textbf{Dataset Availability.}
Related work often does not put restrictions on the dataset available to the attacker, except for limiting the amount of ground-truth labels. 
We find that the attacker's dataset significantly influences the effectiveness of the removal attacks. 
Increasing the amount of (unlabeled) domain data is sufficient to perform successful removal attacks, and predicted labels can substitute ground-truth labels. 

We found that using a transfer dataset (labeled data from a different domain) to train a model from scratch, such as in the Knockoff attack~\cite{orekondy2019knockoff}, does not lead to successful removal attacks.
For CIFAR-10, almost all watermarks are retained, and for ImageNet we could not train a surrogate model with high test accuracy. 
We observe that access to domain data is crucial to perform these attacks. 

\textbf{Availability of Pre-Trained Models.}
Related work has not used transfer learning to remove watermarks, but transfer learning is a known method for training models in the visual domain~\cite{torrey2010transfer}. 
We show that transfer learning is highly effective at removing watermarks; it is computationally efficient, and it can leverage access to less data than other model extraction attacks. 
Related work has shown that access to larger transfer sets can reduce the amount of domain data required for transfer learning~\cite{kolesnikov2019big}. 
Specifically, the authors use models that have been pre-trained on up to 300 million images and show that they can transfer learn this model for ImageNet with a test accuracy of 87.5\% using as few as ten examples per class. 
We argue that it should not be a problem for an attacker to obtain access to a pre-trained model from a different domain in practice.
There exist many platforms to share pre-trained models with various model architectures, such as ONNX\footnote{\href{https://onnx.ai/}{https://onnx.ai/}} or Model Zoo\footnote{\href{https://modelzoo.co/}{https://modelzoo.co/}}, without charging the user. 

\subsection{Guidelines}
\label{sec:guidelines}
In this section, we propose guidelines for evaluating the robustness of watermarking schemes. 
These guidelines incorporate many of our findings and provide a minimal checklist to claim robustness for a watermarking scheme.  

\textbf{Attacker's Dataset.} 
Our experiments have shown that robustness on CIFAR-10 does not imply robustness on ImageNet and vice versa. 
In general, we observed that it is more difficult to remove watermarks from models trained on ImageNet than from models trained on CIFAR-10.
We believe that is because (i) the model and task are more complex and (ii) attacks have a greater impact on the model's utility (measured by the test accuracy).
Our recommendation for image classification models is to experiment on (i) a small dataset, (ii) a dataset with large input image dimensions, and (iii) a dataset with a large number of classes.
We use ImageNet to cover the last two requirements within one dataset. 
Furthermore, we recommend listing the amount of data and ground-truth labels used during the attack for removal attacks. 

\textbf{Decision Threshold.}
We noticed that a method to derive a watermarking scheme's decision threshold is missing from many papers in related work. 
Disproving the robustness claim of a scheme requires a method of deriving the decision threshold.
This method affects the scheme's usability. 
For example, for the watermarking scheme Adi, we could theoretically derive the decision threshold because the input images and target labels are drawn randomly. 
However, Blackmarks requires an empirical method to derive a decision threshold because it relies on adversarial examples for which it is difficult to theoretically quantify the transferability of these examples to unmarked models. 
Our work proposes a general method to empirically determine this decision threshold, which involves training many unmarked models on CIFAR-10 and ImageNet (hence the usability is limited).

\textbf{Parameter Ablation.}
We recommend stating all parameters for a removal attack and watermarking scheme that can be included in an ablation study.
In our paper, we manually selected parameters to include in our ablation study. 
For multiple parameters, the robustness should be evaluated at the Nash equilibrium. 
This enhances (i) reproducibility of robustness claims and (ii) allows for a fair evaluation of a scheme's robustness and an attack's effectiveness.

\textbf{Class Accuracies.}
For some watermarking schemes, such as Content or Jia, we observed that the source model might unlearn a single class during the embedding process. 
On ImageNet, the test accuracy drops only by about 0.1\% when the model unlearns a single class, but we argue that in such cases, the impact of the watermark is greater than the drop in overall test accuracy is suggesting. 
We recommend to evaluate the drop in test accuracy for single classes.

\textbf{Runtime.}
We suggest that a watermarking scheme or removal attack should show their runtimes for the embedding or removal procedure in relation to retraining a model from scratch. 
While the runtime of all surveyed watermarking schemes is small, we believe the runtime is still a distinguishing factor for the proposed scheme's practicality. 

\subsection{Implications for Future Research}
We show with our systematic, empirical study that a well-defined attacker can break all surveyed watermarking schemes. 
We argue that DNN watermarking robustness needs to be defined and evaluated more rigorously. 
Many previous works evaluate against a relatively weak attacker that does not adapt their attacks. In other cases, the attacker is limited by their computational resources or the non-availability of other pre-trained models. 
We present a well-defined attacker model and our Watermark-Robustness-Toolbox\footnote{\url{https://github.com/dnn-security/Watermark-Robustness-Toolbox}} is publicly available. 
Authors of future watermarking schemes can evaluate robustness against the attacker presented in this paper. 
Our paper does not imply that DNN watermarking is impossible and there exist fingerprinting schemes~\cite{lukas2019deep} that show promising results. 

\section{Conclusion}
\label{sec:conclusion}
We have proposed taxonomies for DNN watermarking schemes and removal attacks. 
The taxonomies define four categories of watermarking schemes and three categories of removal attacks. 
We evaluate eleven watermarking schemes from related work and empirically determine their decision thresholds for the CIFAR-10 and ImageNet datasets.
Then, we measured the performance of a large set of removal attacks against all watermarking schemes and ablate over multiple parameters for each scheme and removal attack.
We use the Nash equilibrium to evaluate a scheme's robustness against (i) all attacks, (ii) categories of attacks, and (iii) single attacks.
Our results show that none of the schemes is robust against all attacks.
We break down these results by analyzing each attack category's effectiveness and find that the most effective removal attack category are model extraction attacks, followed by model modification attacks. 
We show that transfer learning removes all watermarks on CIFAR-10, but there exists no such dominant attack for ImageNet. 
We create a combined attack composed of (1) transfer learning and (2) label smoothing that removes all eleven watermarks. 
Finally, we discuss the practicality of the removal attacks, e.g., their monetary costs and the dataset availability of the attacker and propose guidelines for evaluating the robustness of DNN watermarking.  
We hope that our work will improve future evaluations of DNN watermarking schemes.

\bibliographystyle{IEEEtran}
\bibliography{sample.bib}

\section{Appendix}
The Appendix is organized as follows.
Section~\ref{sec:datasets} describes the datasets used in our experiments. 
Section~\ref{sec:schemes} describes all surveyed watermarking schemes and the parameters we used in our ablation study.
Section~\ref{sec:attacks} describes all surveyed removal attacks including novel attacks such as weight shifting and contains a description of the parameters we used in the ablation study.
A detailed description of each approach can be found in the author's papers. 
Section~\ref{tab:decision_threshold} provides further details on the computation of the decision thresholds (see Section~\ref{sec:measurements}).

\section{Watermarking Schemes}
\label{sec:schemes}

In this section, we present the surveyed watermarking schemes and the parameters used for our ablation study.
For simplicity, we refer to a watermarking scheme by the first author's name unless it is known under a different name.

\subsection{Model Independent}
\textbf{Adi}~\cite{adi2018turning}.
We embed the same $100$ watermarking keys used by the authors\footnote{\href{https://github.com/adiyoss/WatermarkNN}{https://github.com/adiyoss/WatermarkNN}}.
Images are resized along their shortest side to the dimensions of the training data, followed by center cropping.  
For ImageNet, we embed using early stopping~\cite{cataltepe1999no} on the watermarking loss with a patience of five, evaluated at the end of every 200th batch.
The watermarking loss is the cross-entropy loss of the model computed on the watermarking key.
We ablate over the learning rate $\text{lr} \in \{10^{-3}, 10^{-4}\}$.
To speed up the embedding, we repeat the watermarking keys 1000 times for ImageNet and 100 times for CIFAR-10.  

\textbf{Zhang}~\cite{zhang2018protecting}. The authors propose three different schemes, referred to as \emph{Content}, \emph{Noise} and \emph{Unrelated}.  
\begin{itemize}
    \item \textbf{Content}:
    We use a white square embedded at the top left corner of the image.
    The square's size is $s\in \{32, 128\}$ for ImageNet and $s\in\{8,16\}$ for CIFAR-10. 
    \item \textbf{Noise}:
    We add the noise across the entire image and clip the resulting values into the range $[0,1]$. 
    We ablate over the standard deviation $\sigma\in \{0.4, 1.0\}$ for both ImageNet and CIFAR-10.
    For CIFAR-10, we ablate over the learning rate during the embedding $\text{lr} \in \{10^{-3}, 10^{-4}\}$. 
    \item \textbf{Unrelated}:
    We sample watermarking images from the Omniglot dataset~\cite{lake2015human} for both CIFAR-10 and ImageNet. 
    We ablate over the learning rate $\text{lr} \in \{10^{-3}, 10^{-4}\}$. 
\end{itemize}
For CIFAR-10, we randomly sample the source-target class pair 'cat' and 'dog' and for ImageNet, we sample 'tiger shark' and 'stingray'. 
We use early stopping on the watermarking loss during the embedding and repeat the watermarking keys 1000 times for ImageNet and 100 times for CIFAR-10. 

\subsection{Model Dependent}
\textbf{Frontier-Stitching}~\cite{le2020adversarial}. We use FGM~\cite{goodfellow2014explaining} to generate adversarial examples and ablate over the perturbation threshold perturbation threshold $0.1 \leq \epsilon \leq 0.25$.

\textbf{Blackmarks}~\cite{chen2019blackmarks}. We ablate over the loss term that minimizes the bit error rate between the predicted cluster and the assigned cluster $0.01 \leq \lambda \leq 100$.

\textbf{Jia}~\cite{jia2020entangled}.
We sample the watermarking key from the training data and use a square as the secret trigger pattern (same as the authors). 
For CIFAR-10, we compute the source class 4 ('deer') and target class 6 ('frog'). 
We use SNNL weights $w\in \{0.25, 1, 4\}$ and a rate of $r=2$, i.e., every second batch consists of watermark data. 
The trigger has a size of $3\times 3$ pixels and resets values to zero in the image across all three channels. 
For ImageNet, we compute source class 3 ('tiger shark, Galeocerdo cuvieri') and target class 4 ('hammerhead, hammerhead shark').
We use an SNNL weight $w=64$ and a ratio of ten during the embedding using a square trigger with $5\times 5$ pixels.
We compute the SNNL on a single layer, as mentioned by the authors, due to GPU memory restrictions when computing the SNNL on all layers. 
When embedding $100$ elements with a batch size of 64, we observe the convergence of the SNNL and cross-entropy losses after about 100k images are shown to the source model. 

\subsection{Parameter Encoding}

\textbf{Uchida}~\cite{uchida2017embedding}.
We embed the Uchida watermark with early stopping on the loss during training and a patience of five, whereby we evaluate the condition at the end of every epoch for CIFAR-10 and after every 200 batches for ImageNet. 
The target layer has $9\,408$ weights for the ImageNet models and $432$ for CIFAR-10 models. 
For CIFAR-10, we ablate over the constant weight factor of the embedding loss $\lambda\in \{0.1, 1, 10\}$ and for ImageNet, we ablate over $\lambda \in \{1, 10\}$.

\textbf{DeepMarks}\footnote{DeepMarks is labeled as a fingerprint by the authors, but since it modifies the model by embedding a message, it is a watermark as per our definition.}~\cite{chen2018deepmarks}. We ablate over the embedding strength $\gamma \in \{0.1, 10\}$ and use the same target layer as in Uchida. 

\textbf{DeepSigns}~\cite{rouhani2018deepsigns}.
In the author's paper, clusters are modelled using a Gaussian Mixture Model, whereby each feature cluster $c_i$ is described by a mean $\mu_i$ and a standard deviation $\sigma_i$. 
In our experiments, we had difficulties embedding the watermark in ImageNet models using more than $m=1$ Gaussian distributions because of instabilities during training. 

Even after extensive parameter search, we observe that for $m>1$ (i) the test accuracy drops significantly over time, and (ii) the regularization loss does not converge. 
The authors do not provide source code, nor did they validate their scheme for ImageNet. 
We solve the issue for ImageNet by modifying two elements of the embedding procedure.
\begin{itemize}
    \item \textbf{Single Gaussian}: We use $m=1$ Gaussian and $n=100$ bits to embed the message on ImageNet. 
    We observe that the regularization loss converges. 
    \item \textbf{Alternating Training}: We train on the whole dataset without the regularization loss for two batches.
    Then, we fine-tune with the embedding loss on samples from the source class for one batch. 
    We observe that this stabilizes training and maintains a high test accuracy.
\end{itemize}
We can replicate the author's result on CIFAR-10 by using $m=10$ Gaussian distributions (one for each class) and embedding $n=10$ bits per Gaussian. 
On ImageNet, we embed the watermark into a layer with $25\,088$ features and  $24\,576$ features for CIFAR-10. 

\subsection{Active Schemes}
\textbf{DAWN}~\cite{szyller2019dawn}. 
We ablate over the expected rate $r\in \{0.01, 0.02\}$ at which a false label is returned. 

\section{Watermark Removal Attacks}
\label{sec:attacks}

In this section, we describe the parameters used in our ablation study for all removal attacks surveyed in this paper, sorted by their attack category.
We make configuration files that show the parameter ablations for all removal attacks publicly available as part of our Watermark-Robustness-Toolbox (WRT).  
A summary of the adversary model for each attack (see Section~\ref{sec:adversary_model}) is listed in Table~\ref{tab:removal_attacks}.

\subsection{Input Preprocessing}
\textbf{Input Reconstruction}~\cite{lin2019invert}. uses an autoencoder\footnote{\href{https://github.com/foamliu/Autoencoder}{https://github.com/foamliu/Autoencoder}}~\cite{ng2011sparse} to compress and reconstruct images before passing them to the surrogate model.
We ablate over the size of its bottleneck layer $64 \leq h \leq 512$. 
We do not perform Input Reconstruction on ImageNet because, to the best of our knowledge, no high-fidelity autoencoder for ImageNet is available. 

\textbf{Input Noising}~\cite{zantedeschi2017efficient}.
We ablate over the standard deviation $0.01 \leq \sigma \leq 0.2$ for Gaussian noise with zero mean. 

\textbf{Input Quantization}~\cite{lin2019defensive}. 
For a given number of bits $b$ we discretize the input space into $2^b$ evenly spaced intervals, referred to as \emph{quanta}. 
We project every value of the input image to the mean of its quantum and ablate over the number of bits $b\in \{3,4,5\}$. 

\textbf{Input Smoothing}~\cite{xu2017feature}.
We use a mean, median, and Gaussian kernel. 
For the mean and median kernels, we use a filter size of three, and for the Gaussian kernel, we ablate over the standard deviation $0.1 \leq \sigma \leq 0.3$. 

\textbf{Input Flipping}. We flip an image along its horizontal axis. 

\textbf{JPEG Compression}~\cite{dziugaite2016study}.
We ablate over a parameter $5 \leq q \leq 95$ that controls the quality of the compression. 

\textbf{Feature Squeezing}~\cite{xu2017feature}. The quanta values are chosen to be multiples of $0.5^k$ for some $1 \leq k \leq 6$.

\subsection{Model Modification}

\textbf{Adversarial Training}~\cite{madry2017towards}.
We inject about 10\% of the training dataset's size with adversarial examples generated using Projected Gradient Descent~\cite{madry2017towards} for $\epsilon \in \{0.01, 0.1, 0.25\}$, a step size of $0.01$ and a maximum number of $40$ iterations. 
Each adversarial example is repeated twice, and we fine-tune the surrogate model for five epochs. 

\textbf{Feature Permutation.} DNNs are invariant to feature permutations, meaning that neurons in a hidden layer can be permuted without affecting the model's functionality. 
We use (random) feature permutation as an adaptive attack designed specifically against Deepsigns~\cite{rouhani2018deepsigns}, which encodes the message into the activations of hidden layers. 

\textbf{Fine-Pruning}~\cite{liu2018fine}.
We ablate over the sparsity $0.8\leq \rho \leq 0.95$ and fine-tune for ten epochs on CIFAR-10 and five epochs on ImageNet. 

\textbf{Fine-Tuning}~\cite{uchida2017embedding}. Fine-Tuning as a model stealing attack refers to a set of attacks that first apply a transformation to the model, followed by fine-tuning. 
\begin{itemize}
    \item Fine-Tune All Layers (\textbf{FTAL}). All weights are fine-tuned. 
    \item Fine-Tune Last Layer (\textbf{FTLL}). All but the last layer's weights are frozen while the model is fine-tuned.
    \item Retrain All Layers (\textbf{RTAL}). The last layer's weights are re-initialized, and all weights are fine-tuned. 
    \item Retrain Last Layer (\textbf{RTLL}). The last layer's weights are re-initialized, and only that layer's weights are fine-tuned. 
\end{itemize}
RTAL and RTLL use predicted labels, whereas FTAL and FTLL use ground-truth labels (otherwise, gradients are zero). 

\textbf{Label Smoothing}~\cite{szegedy2016rethinking}. 
We use a weight of $\epsilon=0.3$ for the weighted sum between the prediction and a uniform vector. 

\textbf{Regularization}~\cite{shafieinejad2019robustness}.
We L2-regularize for five epochs on CIFAR-10 and one epoch on ImageNet using a weight decay of 0.1 (two orders of magnitudes higher than during training). 

\textbf{Neural Cleanse}~\cite{wang2019neural}.
We implement both \textit{unlearning} and \textit{pruning} methods proposed by the authors and ablate over the learning rate $10^{-3} \leq \alpha \leq 10^{-2}$ for unlearning and the sparsity $0.8 \leq \rho \leq 0.99$ for pruning.

\textbf{Neural Laundering}~\cite{aiken2020neural}.
We ablate over the activation threshold to prune convolutional layer neurons $0.03 \leq c \leq 3$ and the learning rate for fine-tuning $10^{-4} \leq \alpha \leq 10^{-2}$.

\textbf{Weight Pruning}~\cite{zhu2017prune}. 
We ablate over the sparsity $0.1 \leq \rho \leq 0.95$ for the trainable weights of each layer.

\textbf{Weight Shifting}. Weight shifting is a novel, adapted attack against parameter encoding watermarking schemes. 
The idea is to apply a small perturbation to all filters of each convolutional layer in the network, followed by fine-tuning the model to regain the loss in test accuracy. 
We design weight shifting as an efficient and effective model stealing attack specifically against Uchida~\cite{uchida2017embedding} and Deepmarks~\cite{chen2018deepmarks}. 

We explain the attack's idea at the example of Uchida, but a similar intuition holds for Deepmarks where the extraction is highly similar. 
Let $W \in \mathbb{R}^{n\times c \times w\times}$ be the convolutional filters of a target layer, where $n$ is the number of filters, $c$ are the number of channels, and $w,h$ are the width and height of each filter. 
A weakness of Uchida exploited by weight shifting is that the attacker knows that if all convolutional filters were inverted, i.e. $W_i' = -W_i$, then the watermark accuracy would be zero. 
We cannot directly invert all filters, as the model experiences a significant drop in test accuracy.
Hence, we construct a 'softer' version of the attack that only moves each filter in the direction of the inverse mean multiplied by some constant weight parameter $\lambda_1\in \mathbb{R}$. 
We additionally add small random Gaussian noise to each filter to encourage the network to find slightly different filters in the fine-tuning phase. 

Our attack can be formalized by the function $S(W; \lambda_1, \lambda_2)$, which takes as input a set of filters $W$ and outputs a shifted set of filters $W'$. 
The parameter $\lambda_1, \lambda_2$ trade off the attack's efficiency with its effectiveness. 
Let $A$ be a random normal matrix of the same shape as each filter $W_i$ with a variance equivalent to the variance over all filters for a convolutional layer and a mean of zero.  
Shifted weights for each convolutional layer can be computed by applying the following function. 
\begin{align}
    \text{S}(W; \lambda_1, \lambda_2)_i = W_i - \frac{\lambda_1}{n} \sum_{j=1..n} W_j - \lambda_2 A
\end{align}
In our experiments, we use $\lambda_1=1.5, \lambda_2=1.0$ for CIFAR-10 and $\lambda_1=1.3, \lambda_2=0$ for Imagenet.
We fine-tune the model for ten epochs on CIFAR-10 and for five epochs on ImageNet. 

\textbf{Weight Quantization}~\cite{hubara2017quantized}.
We ablate over the bit-size $b\in \{4, 5\}$ (i.e., there are $2^b$ discrete states) for CIFAR-10 and ImageNet and fine-tune the model for one epoch.  

\subsection{Model Extraction}
\textbf{Retraining}~\cite{tramer2016stealing}. We use the same parameters for the surrogate model that were used to train the source model. 

\textbf{Smooth Retraining}. Smooth Retraining trains a surrogate model on smoothed labels obtained from querying the source model for multiple variations of the same image. 
For each query, a random, affine transformation (e.g., random cropping) is applied to the image, and the mean of all received labels is computed as the final label.
We design smooth retraining as an adaptive attack against the active watermarking scheme DAWN.
The intuition is that if DAWN responds with a false label for one image, variations of the same image have a high probability of receiving the label predicted by the source model. 
In our experiments, we use $n=3$ queries.
     
\textbf{Knockoff Nets}~\cite{orekondy2019knockoff}.
We implement the random selection approach on the Open Images~\cite{openimages} dataset.

\textbf{Transfer Learning}~\cite{torrey2010transfer}. Transfer Learning is an established method from related work, where a pre-trained model from a different domain is fine-tuned for a new domain.
We propose using transfer learning as a novel method to remove DNN watermarks. 
We use a pre-trained ResNet-101 model\footnote{\href{https://storage.googleapis.com/openimages/2017_07/oidv2-resnet_v1_101.ckpt.tar.gz}{https://storage.googleapis.com/openimages/2017\_07/oidv2-resnet\_v1\_101.ckpt.tar.gz}} for Open Images (v2)~\cite{openimages} that was published by Google in 2017. 
The model defines an output layer with 5k output classes, which we replace by a layer with ten output classes for CIFAR-10 and 1k output classes for ImageNet. 
We transfer-learn the model using stochastic gradient descent (SGD) and freeze all but the last layer for the first 300 batches. 
We proceed by training the entire model for five epochs and reduce the learning rate by a factor of ten in epochs three and four. 

\textbf{Adversarial Training (from scratch)}~\cite{madry2017towards}. This method is equivalent to adversarial training described earlier, except that the attacker trains the surrogate model from scratch. 

\section{Datasets}
\label{sec:datasets}

We now describe the datasets used in our experiments. 

\begin{itemize}
    \item \textbf{CIFAR-10}~\cite{cifar10} contains $50$k training images and $10$k testing images from 10 classes. 
    All images have a resolution of $32 \times 32$ pixels.
    \item \textbf{ImageNet}~\cite{imagenet} contains $1.28$ million training images and $150$k testing images from 1k classes.
    We resize and center crop all images to $224\times 224$ pixels.
    \item \textbf{Open Images}~\cite{openimages} defines $19.794$ classes and contains in total $8.85$ million training images, out of which we use a subset of $1.7$ million images due to storage constraints on our machines. 
    Images can be labeled by multiple classes. We resize and center-crop all images to $224\times 224$ pixels. 
\end{itemize}
All source models are trained on either CIFAR-10 or ImageNet. 
The Open Images dataset is only used in the transfer learning attack. 
We use standard training procedures and data augmentation, such as horizontal flipping, to train models for CIFAR-10 and ImageNet from scratch. 
On CIFAR-10 and ImageNet, the source models achieve a test accuracy of $94.20\%$ and $75.48\%$ respectively. 

\subsection{Dataset Splitting}
We split the whole training dataset into thirds and assign two-thirds to the defender for embedding the watermark. 
For the attacker's training data, we recall from Section~\ref{sec:attacker_capabilities} that we distinguish between the availability of the following three datasets to the attacker. 
\begin{enumerate}
    \item \textbf{Labeled}: Data from the same distribution where a subset of at most a third of the data is labeled.
    \item \textbf{Domain}: Unlabeled data from the same distribution.
    \item \textbf{Transfer}: Labeled data from a different distribution.
\end{enumerate} 
In the first two cases, we assign the remaining third of the training dataset to the attacker. 
We make an exception for model extraction attacks, where the attacker has access to the whole training dataset without labels. 
Such an exception is necessary because model extraction attacks require a substantial amount of data to output well-trained surrogate models. 
We underpin this argument by an ablation study in Section~\ref{sec:exp_robustness}.
Otherwise, if the attacker is given domain data, we replace all labels with the predictions of the source model. 

\section{Estimating the Decision Threshold}
\label{sec:appendix_decision_threshold}

For model independent, model dependent and active watermarking schemes, we use 20 publicly available, pre-trained models from the torchvision\footnote{\href{https://pytorch.org/vision/stable/models.html}{https://pytorch.org/vision/stable/models.html}} package that do not necessarily share the source model's architecture (ResNet-50). 
We use the following model architectures. 

ResNet-18, ResNet-34, ResNet-50, ResNet-101, ResNet-152~\cite{he2016identity}, Wide ResNet-50, Wide ResNet-101~\cite{zagoruyko2016wide}, VGG11, VGG13, VGG16, VGG19~\cite{simonyan2014very}, SqueezeNet~\cite{iandola2016squeezenet}, DenseNet-121, DenseNet-161~\cite{huang2017densely}, 
GoogleNet~\cite{szegedy2015going}, Alexnet, Alexnet-50~\cite{krizhevsky2012imagenet}, InceptionNet~\cite{szegedy2016rethinking}, MobileNetV2~\cite{sandler2018mobilenetv2}

\end{document}